# Kinesin-8 effects on mitotic microtubule dynamics contribute to spindle function in fission yeast

Zachary R. Gergely[1,2], Ammon Crapo[1], Loren E. Hough[1], J. Richard McIntosh[2], and Meredith D. Betterton[1]

*Departments of [1]Physics and [2]MCD Biology, University of Colorado at Boulder*

April 17, 2016

**Abstract**

Kinesin-8 motor proteins destabilize microtubules. Their absence during cell division is associated with disorganized mitotic chromosome movements and chromosome loss. Despite recent work studying effects of kinesin 8s on microtubule dynamics, it remains unclear whether the kinesin-8 mitotic phenotypes are consequences of their effect on microtubule dynamics, their well-established motor activity, or additional unknown functions. To better understand the role of kinesin-8 proteins in mitosis, we have studied the effects of deletion of the fission-yeast kinesin-8 proteins Klp5 and Klp6 on chromosome movements and spindle length dynamics. Aberrant microtubule-driven kinetochore pushing movements and tripolar mitotic spindles occurred in cells lacking Klp5 but not Klp6. Kinesin-8 deletion strains showed large fluctuations in metaphase spindle length, suggesting a disruption of spindle length stabilization. Comparison of our results from light microscopy with a mathematical model suggests that kinesin-8 induced effects on microtubule dynamics, kinetochore attachment stability, and sliding force in the spindle can explain the aberrant chromosome movements and spindle length fluctuations seen.

## 1 Introduction

Kinesin-8 proteins are motor enzymes that can alter microtubule dynamics (Messin and Millar, 2014). Members of the kinesin-8 family include Kip3 in budding yeast (DeZwaan et al., 1997), Klp5 and Klp6 in the fission yeast *S. pombe* (West et al., 2001; Garcia et al., 2002b), KLP67A in Drosophila (Pereira et al., 1997), and Kif18A (Zhu and Jiang, 2005), Kif18B (Stout et al., 2011), and Kif19 (Tanenbaum et al., 2009) in humans. Kinesin 8s across organisms are motors that walk toward the plus ends of microtubules (Pereira et al., 1997; Gupta et al., 2006; Mayr et al., 2007; Grissom et al., 2009; Niwa et al., 2012) with high processivity (Varga et al., 2006, 2009; Jannasch et al., 2013). Upon reaching the microtubule (MT) end they promote its destabilization (DeZwaan et al., 1997; West et al., 2001; Gupta et al., 2006; Varga et al., 2006; Unsworth et al., 2008; Tischer et al., 2009; Gardner et al., 2011). Both in cells and in reconstituted systems, the directed motility of kinesin 8s typically leads to an increase in their concentration near MT plus ends (Gupta et al., 2006; Varga et al., 2006; Mayr et al., 2007; Stumpff et al., 2008; Tischer et al., 2009; Masuda et al., 2011; Fukuda et al., 2014), an effect enhanced by an additional MT binding site in the tail of the motor (Mayr et al., 2011; Stumpff et al., 2011; Su et al., 2011; Weaver et al., 2011; Erent et al., 2012). As a result, kinesin-8 motors can increase MT catastrophe frequency preferentially for longer MTs (Gupta et al., 2006; Tischer et al., 2009; Gardner et al., 2011; Stumpff et al., 2011; Fukuda et al., 2014), an effect which can contribute to MT length regulation (Varga et al., 2006, 2009; Hough et al., 2009; Kuan and Betterton, 2013). While Kip3 can depolymerize chemically-stabilized MTs (Gupta et al., 2006; Varga et al., 2006, 2009; Su et al., 2011), this activity has not been observed for the *S. pombe* kinesin 8 (Grissom et al., 2009;

Erent et al., 2012), and for Kif18A different results have been reported by different groups (Mayr et al., 2007; Du et al., 2010).

Kinesin-8 motors are typically considered to promote MT depolymerization, but they have been reported to affect surprisingly many aspects of MT dynamics and interactions, including growth speed (Tytell and Sorger, 2006; Unsworth et al., 2008; Fukuda et al., 2014), shortening speed (Tytell and Sorger, 2006; Grissom et al., 2009; Su et al., 2011; Fukuda et al., 2014), rescue frequency (Gupta et al., 2006; Unsworth et al., 2008; Fukuda et al., 2014), pausing (Stumpff et al., 2012), nucleation (West and McIntosh, 2008; Erent et al., 2012), tubulin dimer binding (Erent et al., 2012), tubulin ring formation (Peters et al., 2010), MT crosslinking (West and McIntosh, 2008; Grissom et al., 2009; Erent et al., 2012; Su et al., 2013), and sliding (Su et al., 2013). Some work has proposed that kinesin 8s act as dampeners of MT dynamics, increasing both catastrophe and rescue frequency to increase MT dynamicity (Stumpff et al., 2008; Unsworth et al., 2008; Du et al., 2010; Stumpff et al., 2011; Su et al., 2011; Stumpff et al., 2012; Kim et al., 2014; Fukuda et al., 2014), an effect that may occur due to MT-binding and -stabilizing activity of the tail (Su et al., 2011, 2013; Fukuda et al., 2014). In budding yeast, the catastrophe-promoting versus rescue-promoting activities of Kip3 appear to be spatially regulated (Fukuda et al., 2014).

Studies in which kinesin-8 motors are altered or deleted in living cells show that they play some roles in mitosis, but precisely how they contribute to organized chromosome movements and mitotic spindle function is not completely clear. Deletion or knockdown of kinesin-8 proteins leads to longer mitotic spindles in many cases (Cottingham and Hoyt, 1997; Straight et al., 1998; West et al., 2002; Goshima and Vale, 2003; Gandhi et al., 2004; Mayr et al., 2007; Stumpff et al., 2008; Wang et al., 2010) but not always (Wargacki et al., 2010; Su et al., 2013). Kinesin-8 deletion or depletion is associated with defects in spindle assembly (Gandhi et al., 2004; Gatt et al., 2005; Du et al., 2010; Su et al., 2013), positioning (DeZwaan et al., 1997; Cottingham and Hoyt, 1997; Fukuda et al., 2014), and elongation (Gatt et al., 2005; Wang et al., 2010). Kinesin 8s localize to kinetochores (KCs) (West et al., 2002; Garcia et al., 2002a; Tytell and Sorger, 2006) and kinetochore MTs (Mayr et al., 2007; Stumpff et al., 2008; Masuda et al., 2011; Kim et al., 2014). Kinesin 8 deletion or knockdown causes multiple chromosome movement defects including alterations in chromosome congression and larger chromosome oscillations (West et al., 2002; Garcia et al., 2002a; Sanchez-Perez et al., 2005; Mayr et al., 2007; Stumpff et al., 2008; Wang et al., 2010; Wargacki et al., 2010; Stumpff et al., 2011), lagging chromosomes (West et al., 2002; Garcia et al., 2002a; Tytell and Sorger, 2006; Stumpff et al., 2008), disruption of spindle-KC attachment (Garcia et al., 2002a; Sanchez-Perez et al., 2005), and an increase in chromosome loss in mitosis (Garcia et al., 2002a; Gandhi et al., 2004).

Kinesin 8s play a role in spindle assembly checkpoint silencing; the checkpoint remains active in kinesin-8 deletion/knockdown (Garcia et al., 2002a; Mayr et al., 2007; Meadows et al., 2011), causing delays in anaphase onset (West et al., 2002; Garcia et al., 2002b,a; Mayr et al., 2007). Both human and fission-yeast kinesin 8s bind to protein phosphatase 1 (PP1), an interaction that helps localize PP1 at KCs and silence the checkpoint (Meadows et al., 2011; Tang et al., 2014; De Wever et al., 2014). Consistent with a decrease in depolymerizing activity in kinesin-8 depletion/knockdown, most previous work has observed a decrease in bioriented sister KC separation, suggesting a decrease in KC tension (Garcia et al., 2002a; Zhu and Jiang, 2005; Mayr et al., 2007; Stumpff et al., 2008; Wargacki et al., 2010); however, some work has reported an increase (West et al., 2002; Tytell and Sorger, 2006).

Kinesin-8 deletion/knockdown can alter the speed of chromosome movements in anaphase (Stumpff et al., 2008; Tang et al., 2014). Kinesin-8 proteins localize to the anaphase spindle midzone (West et al., 2002; Garcia et al., 2002a) and their depletion is associated with destabilization of the interzone (Gandhi et al., 2004; Gatt et al., 2005), altered sliding (Wang et al., 2010; Su et al., 2013), and longer anaphase spindles (Rizk et al., 2014). Kinesin-8 deletion mutants can also exhibit defects in spindle disassembly (Woodruff et al., 2010).

The variability in reported roles of kinesin 8s suggest that their effects may be different in different organisms. In fission yeast, Klp5/6 form a heterodimer (Garcia et al., 2002b) and deletion of either protein has shown similar phenotypes during vegetative growth (West et al., 2002; Unsworth et al., 2008). Klp5 alone can form a homodimer, while Klp6 cannot (Garcia et al., 2002b; Li and Chang, 2003). On interphase MTs, Klp5/6 increases MT growth speed and catastrophe and rescue frequencies (Unsworth et al., 2008) and can cause a MT-length-dependent increase in catastrophe frequency (Tischer et al., 2009), an activity that contributes to nuclear centering (Glunˇciˊc et al., 2015). Purified Klp5/6 can track depolymerizing MT ends, crosslink and bundle MTs, and bind soluble tubulin dimers (Grissom et al., 2009; Erent et al., 2012). Deletions of Klp5/6 lead to longer mitotic spindles, increased movements of chromosomes back and forth along the spindle, a delay in anaphase onset, destabilized spindle-KC attachments, and an increase in chromosome loss in mitosis (West et al., 2001, 2002; Garcia et al., 2002a,b). Localization of Klp5/6 at KCs is promoted by binding interactions with Alp7/TACC, which forms a complex with the Ndc80 internal loop (Tang et al., 2014). Some previous work has found mitotic nuclear localization of Klp5 or 6 is retained if the partner gene is deleted (West et al., 2001), while others have found interdependent localization (Unsworth et al., 2008).

While much recent work has focused on effects of kinesin 8s on MT dynamics, we do not yet understand the extent to which the kinesin-8 mitotic phenotypes are consequences of their effect on MT dynamics, their well-established motor activity, or some additional function(s). To better understand the roles of kinesin-8 proteins in mitosis, we used quantitative fluorescence microscopy and mathematical modeling to study the effects of their deletion on chromosome movements and spindle length dynamics in fission yeast.

## 2 Experimental Results

Fission yeast strains carrying a cold-sensitive *β*-tubulin, *nda3-KM311*, experience MT depolymerization at 18°C. Cold treatment arrests these cells in early mitosis (Hiraoka et al., 1984). During this arrest, chromosomes often lose their interphase attachment to the site on the nuclear envelope to which the spindle pole body (SPB) is attached (Grishchuk and McIntosh, 2006), so they become free in the nucleoplasm or "lost" from normal SPB association. After shift to permissive temperature, MTs polymerize, the spindle assembles, and lost KCs attach to nuclear MTs emanating from the SPBs (polar MTs) and cells proceed through mitosis (fig. 1A-F, supplemental movies 1-3). Previous work has used this system to observe lost KCs in early mitosis and the reeling in of the reattached KC to the SPB (Grishchuk and McIntosh, 2006).

We constructed cells carrying the *nda3-KM311* allele, SPBs labeled with *pcp1-GFP*, and the centromere of chromosome 2 labeled with *cen2-GFP* (Yamamoto and Hiraoka, 2003). This was our wild type strain to which we added deletions of *klp5*, *klp6* or both (fig. 1A); we refer to the deletion strains as 5Δ6+, 5+6Δ, and 5Δ6Δ. To image MTs, we have constructed strains that additionally contain the *mCherry-atb2* *α*-tubulin allele expressed under a weak promoter (Yamagishi et al., 2012), allowing us to observe the long mitotic spindles characteristic of kinesin-8 deletion (fig. 1B).

To determine whether the multiple fluorescent tags, kinesin-8 deletion, and cold treatment interact in a way that could adversely affect the cells, we measured doubling time before and after cold treatment (table 1). While cold treatment increased the subsequent doubling time, both wildtype and kinesin-8 delete cells showed similar changes in growth rate after cold block. This suggests that cold treatment does not have disproportionate effects in the kinesin-8 deletion background.

We verified by fixed-cell imaging that chromosomes were lost during cold treatment and that KCs became associated with mitotic polar MT bundles immediately after rewarming (fig. 1C), as observed previously (Kalinina et al., 2012). We use the term polar MTs for single or bundled nuclear MTs emanating from an SPB, and the term interpolar MTs for the shaft of interdigitating polar MTs that runs between the poles; in some studies the latter have been called the central or core spindle. (Note that in anaphase, cytoplasmic astral MTs develop; these have not been measured in our study.) For live-cell imaging, we used a temperature-control system that allowed initial imaging at 18°C, where MTs were depolymerized and KCs were frequently lost (fig. 1C-F). We shifted the temperature to 32°C within 1.5 minutes while imaging to quantify the time of MT repolymerization. We were able to routinely capture images in multiple focal planes, imaging the entire volume of the nucleus every 15 s for >30 minutes. We then tracked the locations of fluorescent peaks to determine three-dimensional KC-SPB and SPB-SPB separations as a function of time (Methods).

In wild-type cells, lost KCs typically were reeled in to the SPB, bioriented, and segregated (fig. 1D, supplemental movie 1). In cells with kinesin-8 deletion mutants, we observed aberrant chromosome movements and spindle dynamics, including spindle length fluctuations, lagging chromosomes, and chromosome pushing movements away from the SPB (fig. 1E-F, supplemental movies 2-3).

## 2.1 Kinesin-8 deletion altered chromosome movements after microtubule repolymerization

We first studied the attachment of lost KCs to MTs and subsequent chromosome movements to determine what roles kinesin 8s play in the recapture of lost KCs. We observed three characteristic behaviors: reeling in of the KC to the SPB (fig. 2A-C, supplemental movie 4), pushing movements of the KC away from the SPB (fig. 2D-F, supplemental movies 5-6), and KC hovering near the SPB (fig. 2G-H, supplemental movie 7).

Rapid reeling in of the KC to the SPB occurred for all cell types and led to colocalization of KC and SPB fluorescence (fig. 2A-C, supplemental movie 4). In 5Δ6+ and 5Δ6Δ cells, KC pushing movements away from the SPB to which it was attached occurred (fig. 2D-F, supplemental movies 5-6). We note that the KC movements away from SPBs occurred in 5Δ6Δ cells, demonstrating that these movements were not dependent on kinesin-8 activity. We verified that both reeling and pushing movements were associated with MTs using strains with mCherry-tagged MTs: we observed the GFP KC marker near polar MTs moving both toward and away from the SPB (fig. 2C,F, supplemental movies 4 and 6). Our results are consistent with previous observations of KC interactions either with the plus end or the lateral wall of capturing MTs (Kalinina et al., 2012). Using the soluble nuclear GFP which was visible from our GFP-lac array-based KC marker, we determined that long polar MTs distorted the nuclear envelope without any apparent breakage of the membrane (fig. 2I).

In all kinesin-8 deletion mutants we observed KC hovering, when the KC moved toward the SPB in an apparent reeling-in movement, but then approximately 0.5 $\mu$m from the SPB it slowed and underwent small fluctuations toward and away from the SPB (fig. 2G-H, supplemental movie 7). We did observe subsequent biorientation of hovering KCs and progression to metaphase, which suggested that such KCs have not lost all attachment to MTs.

To quantify the occurrence of aberrant chromosome movements, we examined populations of 20-30 cells per genotype for 20 minutes after temperature shift. In each cell, we determined whether KC pushing or hovering ever occurred (fig. 2J, no pushing or hovering was observed for wild-type cells).

We observed pushing in 20-50% of cells lacking *klp5*; using the Pearson chi-square test for proportions the populations with and without *klp5* are different with $p = 2.3\times10^{-6}$. We observed hovering in about 30% of all kinesin-8 deletion mutant cells; using the Pearson chi-square test for proportions the wild-type and mutant populations are different with $p = 4.7 \times 10^{-4}$. In the same cell populations, we recorded whether KC reeling in to the SPB had occurred at 5-minute intervals from 0 to 20 minutes after temperature shift (fig. 2K). During initial imaging of cells at 18°C, we observed a lost KC in approximately 30-50% of cells. For 5Δ6+ and 5Δ6Δ cells, a larger initial fraction of uncaptured KCs was visible compared to wild type and 5+6Δ. Previous work found that KC-MT attachment occurs approximately exponentially in time (Kalinina et al., 2012). Consistent with this, we fit the data of fig. 2K to an exponential approach to 1 to determine the mean time for KCs to attach and reel in to the SPB. The exponential function fit well, and the reeling time was almost a factor of three shorter in 5+6Δ cells compared to wild-type, but reeling in took a similar or longer time in 5Δ6+ and 5Δ6Δ cells compared to wild type (fig. 2M).

To better understand the connection between polar MT lengths and the chromosome movements we observed, we measured lengths of mitotic polar MTs (fig. 2L), the three-dimensional distance between a KC and the SPB to which it was reeled in (fig. 2N), and speeds of reeling movements (fig. 2O). A lost KC underwent Brownian motion before attaching to a MT. In wild-type and 5+6Δ cells, KC-MT attachment was typically followed by rapid reeling in to the SPB (fig. 2N). In 5Δ6+ and 5Δ6Δ cells, pushing and reeling movements both occurred, sometimes resulting in multiple back-and-forth movements.

The differences in time to reel in to the SPB for wild-type and 5+6Δ cells were consistent with the longer polar MT lengths in kinesin-8 deletion mutants (fig. 2L). In wild-type cells, exponential fits to the polar MT length distribution gave a characteristic length of 0.45 $\mu$m, consistent with previous observations (Kalinina et al., 2012). Kinesin-8 deletion mutants not only contained longer spindles than wild type, as observed previously (West et al., 2002), their polar MTs were 2-3 times longer than in wild-type cells (1.04 $\mu$m for 5+6Δ and 1.26 $\mu$m for 5Δ6+). Kalinina et al. found that MT rotational diffusion is an important contributor to KC capture in fission yeast and rotational diffusion is sensitive to MT length (Kalinina et al., 2012). Their model predicted that when KC attachment occurs by a fixed-length polar MT, doubling the MT length resulted in approximately 3 times fewer lost KCs remaining after 10 minutes. Therefore, the longer polar MTs are a plausible explanation for the faster KC capture observed in our experiments on 5+6Δ strains. Although the 5Δ6+ and 5Δ6Δ strains also had longer polar MTs, their mean reeling in time was longer than in wild type. This presumably occurred due to pushing movements that delayed reeling in after MT-KC attachment. Consistent with this, the mean time to reel in was longer in the 5Δ6Δ strain in which we observed a higher fraction of pushing events (fig. 2M).

The pushing movements sometimes occurred over long distances (up to 6 $\mu$m), and with speeds of approximately 0.5 $\mu$m/min, consistent with the pushing motion being driven by MT polymerization. When pushing moved the KC all the way to the edge of the cell, we sometimes observed reorientation of the spindle, suggesting that the pushing MTs made mechanical connection to the SPB and were capable of generating significant forces (supplemental movie 8). In cells in which pushing motions occurred, we never clearly observed subsequent KC biorientation and segregation; the cells ultimately went in to anaphase without biorientation. This suggests that the aberrant pushing movements could contribute to chromosome loss and aneuploidy in 5Δ6+ and 5Δ6Δ cells.

Because reeling in of KCs toward the SPB can be driven by MT depolymerization and kinesin8 proteins promote depolymerization, we examined the speeds of reeling movements to determine whether kinesin-8 deletion alters their speed (fig. 2O). Using one-way analysis of variance, we found that the speed measurements are significantly different between the different strains ($p = 2.6×10^{-5}$). We used the two-sample t-test to compare speed measurements for each pair of strains, and found strong statistically significant differences for wild-type versus 5Δ6+ ($p = 4 × 10^{-4}$) and 5+6Δ versus 5Δ6+ ($p = 2.6 × 10^{-5}$), and weaker but significant differences for 5+6Δ versus 5Δ6Δ ($p = 1.4 × 10^{-2}$), and 5Δ6+ versus 5Δ6Δ ($p = 3.6 × 10^{-2}$). These results suggest both that kinesin-8 deletion can alter the speeds of reeling movements and that different types of kinesin-8 deletion lead to different speeds of reeling movements.

## 2.2   Klp5 null strains occasionally displayed tripolar mitotic spindles

Our experimental results showing differences in chromosome movements in 5Δ6+ versus 5+6Δ were surprising, because previous work has found similar mitotic phenotypes for deletion of either Klp5 or 6 (West et al., 2002; Unsworth et al., 2008). To rule out the possibility that our strain construction or choice of fluorescent tags might have produced artifacts in our measurements, we constructed cells with a different set of fluorescent tags on the SPBs and KCs. These fission yeast carried the *nda3-KM311* cold-sensitive tubulin, low-level MT labeling with *mCherry-atb2* under a weak promoter, SPBs labeled with *sid4-mCherry*, and KCs labeled with *mis6-GFP* and *mis12GFP*. We studied cells tagged in this way with *klp5* and *klp6* present (wild type), *klp5* deleted (5Δ6+), and *klp6* deleted (5+6Δ).

After cold treatment and subsequent rewarming on the microscope, these cells showed similar phenotypes to those observed with our original tagging strategy. Cold treatment frequently led to lost chromosomes, which were recaptured to allow mitosis to proceed. Spindle length instability occurred in kinesin-8 deletion mutants, but not in wild-type cells. We observed aberrant chromosome pushing movements in 5Δ6+ cells (fig. 3A, supplemental movies 9 and 10) but not in wild-type or 5+6Δ cells. This confirmed that our results were not a tagging artifact.

In addition, because this set of experiments used mCherry for SPB labeling and GFP for KC labeling, we were able to observe an additional *klp5* deletion phenotype. In some 5Δ6+ cells, we observed three SPBs and/or tripolar mitotic spindles (fig. 3B,C, supplemental movies 11 and 12). To examine whether this phenotype was unique to 5Δ6+ cells, we observed samples of cells for up to 30 minutes following shift to permissive temperature. Three SPBs or tripolar spindles were found in 4 out of 108 5Δ6+ cells, a frequency of 3.7%. No cells with three SPBs or tripolar spindles were observed out of 69 wild-type cells and 103 5+6Δ cells. Using the Pearson chi-square test for proportions, the 5Δ6+ and pooled wild-type and 5+6Δ tests are different with $p = 0.011$.

When we observed three SPBs in a single cell, one of the three was typically dimmer than the other two. In some cases we saw what appeared to be a tripolar spindle formed by the splitting of one SPB into two (fig. 3C, supplemental movie 12). This suggested the possibility that the highly stable MTs in 5Δ6+ cells could exert forces large enough to break SPB into two, consistent with the polar-MT driven spindle reorientation we observed in 5Δ6+ cells (supplemental movie 8).

Another possible explanation for our tripolar spindle phenotype could be that chromosome segregation errors lead to changes in numbers of chromosomes and SPBs in the 5Δ6+ cells. To test this possibility, we first tested whether our 5Δ6+ cells had formed stable diploids both by phloxin B staining and attempting to induce sporulation (see Materials and Methods). Both experiments indicated that all of the six 5Δ6+ strains are not stable diploids (data not shown). However, individual cells that we imaged could still have contained altered numbers of KCs and SPBs. To test this possibility, we collected time-lapse images of 5Δ6+ cells, including those of fig. 3, and grouped them into cells that showed apparently normal KC dynamics versus those with aberrant dynamics. We then quantified the total KC brightness in each frame by fitting 3D Gaussian intensity distributions to each spot in our images (Thomann et al., 2002) and using tracking algorithms to identify persistent fluorescent spots (Jaqaman et al., 2008). While the distributions of total KC intensity for the two types of dynamics have similar means (fig. 3D), cells with aberrant dynamics show greater numbers of images with much brighter and dimmer than average total KC intensity. These two intensity distributions are different by the two-sample Kolmogorov-Smirnov test with $p = 1.4 \times 10^{-5}$. Therefore, the tripolar spindle phenotype we observed could have occurred as a result of aneuploidy.

The observation of tripolar mitotic spindles made us question whether the KC pushing phenotype was due solely to tripolar spindle formation. We believe that it was not, both because the tripolar spindle phenotype was so rare, and because chromosome pushing events occurred distinct from tripolar spindles (fig. 3A,C, supplemental movies 9 and 10).

### 2.3    Klp5 and 6 exhibited weak nuclear localization when their partner was absent

The large differences in chromosome movements we observed between $5^+6\Delta$ and $5\Delta$ cells was surprising, because previous work has found similar mitotic phenotypes for deletion of either gene and the double deletion (West et al., 2002; Unsworth et al., 2008). To understand possible origins of these differences, we re-examined the mitotic localization of Klp5/6. In some previous work, deletion of the partner motor gene abolished nuclear localization of Klp5/6-GFP (Unsworth et al., 2008), suggesting that neither Klp5 nor 6 was present in the nucleus of mitotic $5^+6\Delta$, $5\Delta6^+$, and $5\Delta6\Delta$ cells, while other work did not see this (West et al., 2001). Using strains containing GFP-tagged Klp5/6 and mCherry-tagged $\alpha$-tubulin (denoted $5^{GFP}6^+$, $5^{GFP}6\Delta$, $5^+6^{GFP}$, and $5\Delta6^{GFP}$), we imaged both methanol-fixed (fig. 4A) and live (fig. 4B-F) cells using identical camera exposures for each cell. We prepared the images of (fig. 4A-B) with identical brightness and contrast settings for the GFP channel to compare the GFP signal across strains.

In $5^{GFP}6^+$ and $5^+6^{GFP}$ cells, GFP fluorescence was present on interphase MTs, the spindle, and on or near the ends of mitotic polar MTs, consistent with previous work (West et al., 2001; Unsworth et al., 2008). In $5^{GFP}6\Delta$ and $5\Delta6^{GFP}$ cells, GFP signal on the spindle was reduced but still visible. To better view this dim signal, we collected additional images in the $5^{GFP}6\Delta$ and $5\Delta6^{GFP}$ cells with a new EMCCD camera, and processed the images by including fewer focal planes in the maximum-intensity projections and adjusting the brightness and contrast of the GFP signal (fig. 4C-D). These images showed consistent mitotic spindle, near-SPB, and polar MT localization of Klp5/6-GFP in the absence of the partner motor. We note that the near-SPB localization is likely due to Klp5/6 binding to short polar MTs near the SPB, as occurs for Ase1 (Yamashita et al., 2005). The dim Klp5/6-GFP signal may have been detectable in our experiments but not in previous work due to improvements in microscope and camera technology.

Since previous work found that Klp5/6 is exported from the nucleus in the absence of the partner (Unsworth et al., 2008), we considered the possibility that Klp5/6 localization when Klp6/5 is absent could be a transient effect. Two lines of evidence suggest the localization was not transient in our experiments. First, we observed Klp5/6 localization to the spindle in the absence of the partner motor for spindles of a range of lengths, suggesting that the localization persisted throughout mitosis (fig. 4C-D). Second, we collected pairs of images showing that the nuclear localization persisted over time. These images were challenging to collect, because the dim signals in $5^{GFP}6\Delta$ and $5\Delta6^{GFP}$ cells bleached rapidly. Nevertheless, localization could be observed on the same spindle for images collected 5 minutes apart (fig. 4E-F).

We quantified Klp5/6 localization in samples of 50 cells (fig. 4G). In $5^{GFP}6\Delta$ cells, we observed spindle localization in 100% of cells and polar-MT localization on 94% of polar MTs (out of 17 cells with observable polar MTs). We observed near-SPB localization in 12% of cells; note that the consistent spindle localization makes distinct near-SPB localization difficult or impossible to identify. In 5Δ6$^{GFP}$ cells, the localization was dimmer, less consistent, and more patchy, but we observed spindle localization in 34% of cells and 27% of polar MTs (out of 15 cells with observable polar MTs). Near-SPB localization was observed in 62% of cells. (We note that in anaphase, cytoplasmic astral MTs develop that may appear similar to the nuclear polar MTs. However, these anaphase astral MTs can be distinguished from nuclear polar MTs by the length of the spindle at the time of their formation, as in the bottom panel of fig. 4D. We did not measure localization to astral MTs in this study.) The differences in localization between 5$^{GFP}$6Δ and 5Δ6$^{GFP}$ cells were statistically significant (using the Pearson chi-square test for proportions, the $p$ values for 5$^{GFP}$6Δ versus 5Δ6$^{GFP}$ were 2.3 × 10$^{-12}$ for spindle localization, 2.2 × 10$^{-7}$ for near-SPB localization, and 8.4 × 10$^{-5}$ for polar MT localization). The stronger, more consistent localization of 5$^{GFP}$6Δ versus 5Δ6$^{GFP}$ is consistent with the hypothesis that Klp5 can form homodimers but Klp6 remains monomeric.

## 2.4 Biorientation and kinetochore centering on the spindle was disrupted in kinesin-8 deletion cells

We next examined what effect induced KC loss and kinesin-8 deletion had on KC biorientation and centering on the mitotic spindle. We were often able to observe biorientation of KCs after their reeling in to the capturing SPB in wild-type and 5$^+$6Δ cells (fig. 5A, B, supplemental movies 13-14). In 5Δ6$^+$ and 5Δ6Δ, biorientation was rarely observable, probably because the frequently observed pushing movements delayed biorientation and the fact that the cen2 signal in these cells was dimmer (see Materials and Methods). We defined the onset of biorientation to occur when a KC that had been reeled in to (or near) the SPB began to move along the spindle axis between the SPBs. In populations of 20-30 cells per genotype, we observed whether or not KC biorientation had occurred at 5-minute intervals from 0 to 20 minutes after temperature shift and fit the data to an exponential approach to 1 to determine the mean time for KCs to biorient (fig. 5C). The biorientation time was similar for wild-type and 5$^+$6Δ cells, but so much slower in 5Δ6$^+$ and 5Δ6Δ cells that it could not be determined from the fit (fig. 5D). This illustrates the severe biorientation defects in 5Δ6$^+$ and 5Δ6Δ cells: even after KC recapture to the SPB, these mutants showed delays and defects in establishing chromosome biorientation. This may be an important cause of chromosome loss in these cells.

In cells for which biorientation occurred, we measured the separation of the two KC markers (fig. 5E,H). In 5$^+$6Δ cells the mean KC separation was slightly lower than for wild-type cells (422 measurements from 8 cells for wild type and 444 measurements from 8 cells for 5$^+$6Δ). This result is consistent with previous measurements (Garcia et al., 2002a; Zhu and Jiang, 2005; Mayr et al., 2007; Stumpff et al., 2008), but different from earlier observations from our lab, which found larger sister KC separations in kinesin-8 deletion mutants (West et al., 2002). The difference from previous results could be due to the different fluorescent marker used here, differences in MT-KC attachment following cold treatment, or differences in quantitative versus qualitative observations. The duration of biorientation was similar in both strains, 12.9±3.3 min for wild-type (8 cells) and 13.6±2.6 min for 5$^+$6Δ (8 cells).

Kinesin-8 deletion mutants show defects in metaphase chromosome congression and oscillations (West et al., 2002; Garcia et al., 2002a; Sanchez-Perez et al., 2005; Mayr et al., 2007; Stumpff et al., 2008; Wang et al., 2010; Wargacki et al., 2010; Stumpff et al., 2011). Consistent with this, we observed that the distance between the KC and the reeling SPB was different in wild-type and 5$^+$6Δ cells (fig. 5F,G, 422 measurements from 8 cells for wild-type and 444 measurements from 8 cells for 5$^+$6Δ). While wild-type cells showed remarkable centering between the SPBs, the KCs in 5$^+$6Δ cells lost centering and remained closer to the reeling SPB when fractional position along the spindle was measured (fig. 5G); however, because the spindles were larger in 5⁺6Δ cells, the absolute distance from the reeling SPB increased (fig. 5F).

## 2.5 Kinesin-8 deletion led to metaphase spindle length fluctuations

During our imaging experiments, we made the surprising observation that spindle length in kinesin-8 deletion mutants sometimes underwent large fluctuations over time including both elongation and shortening (fig. 1E,F). This was in contrast to wild-type cells and most other organisms, for which spindle length is approximately constant in time. Previous work has shown that kinesin-8 deletion is associated with longer spindles both in fission yeast (West et al., 2002) and in other organisms (Cottingham and Hoyt, 1997; Straight et al., 1998; Goshima and Vale, 2003; Gandhi et al., 2004; Stumpff et al., 2008; Wang et al., 2010). Transient spindle shortening was also reported in fission-yeast kinesin-8 deletion mutants (Syrovatkina et al., 2013). We quantified spindle length by measuring the 3D SPB separation (fig. 6A,B). The dynamics of spindle length in the mutant cells were variable, with some cells showing relatively constant metaphase spindle length, some showing only elongation, and approximately 20% of kinesin-8 deletion cells showing significant spindle length fluctuations.

To determine whether the spindle length fluctuations we observed could have been a type of aberrant anaphase spindle elongation, we compared all wild-type and 5⁺6Δ mutant cells for which the KC marker was clearly visible throughout the experiment. We defined the onset of anaphase by the separation of the sister KCs (fig. 6C). In wild-type cells, significant spindle elongation occurred only after sister KC separation. However, in 5⁺6Δ cells, spindle length fluctuations, including both elongation and shortening, occurred prior to sister KC separation. Therefore, it appears that kinesin-8 deletion caused fluctuations of metaphase spindle length.

We considered whether the temperature-induced depolymerization of MTs followed by repolymerization after temperature shift could lead to artificial spindle length dynamics. However, this possibility seems unlikely, because wild type and kinesin-8 deleted cells showed similar growth rates after cold block (table 1). In some cases, SPBs were already separated at cold temperature on the microscope, perhaps because the spindle had already formed prior to cold block, or due to SPB diffusion during cold treatment. To determine whether initial SPB separation could artificially lead to spindle length fluctuations, we examined cells in which the SPBs were initially not separated. The same spindle length fluctuations occurred in kinesin-8 deletion mutants (data not shown).

To further quantify spindle length dynamics, we measured the temporal autocorrelation function of fluctuations in spindle length for wild-type and 5⁺6Δ spindles (for which metaphase could clearly be identified as the time prior to sister KC separation). The autocorrelation function quantifies how the fluctuation in spindle length at a given time is correlated with the fluctuation at a later time, as a function of the time delay between the two points (see Materials and Methods). When the autocorrelation is zero, the length fluctuations for that time delay have no measurable relationship, while a positive autocorrelation indicates correlated movements and a negative autocorrelation indicates an anticorrelation (for example, if a positive fluctuation at the initial time point is associated with a negative fluctuation at a later time point). The spindle length fluctuation autocorrelation functions were initially positive and decayed to zero after a time delay that indicated the persistence time of the fluctuations. For wild-type cells, the autocorrelation was relatively small, showed a small oscillation to negative values, and decayed to zero by 20 minutes (fig. 6D). For 5⁺6Δ cells, the correlation was larger and underwent a significant oscillation, only reaching zero after 40 minutes (fig. 6D). This indicates persistent dynamics that maintained correlations for the length of our experiments.

We collected all examples of metaphase spindles and performed an automated analysis of spindle length and lengthening and shortening events, for 16-33 cells per strain (see Materials and Methods, table 2). The mean spindle length was longer in kinesin-8 deletion mutants, consistent with previous measurements (West et al., 2002; Syrovatkina et al., 2013). Significant spindle length decreases of several microns, up to 75% of the initial spindle length, occurred in kinesin-8 deletion mutants. The amount and duration of length changes were 2-3 times larger in kinesin-8 deletion mutants than in wild type; the speeds were similar but slightly higher in the deletion mutants.

### 2.6 Kinesin-8 deletion disrupted chromosome segregation

Next we examined the effects of kinesin-8 deletion on movements during chromosome segregation and the timing of segregation. Consistent with previous observations (West et al., 2002; Garcia et al., 2002a; Stumpff et al., 2008; Tang et al., 2014), chromosomes frequently lagged during anaphase in kinesin-8 deletion mutants (fig. 7A,B, supplemental movie 15). In the populations of 20-30 cells per genotype observed for 20 minutes after temperature shift, we observed lagging chromosomes in 10-15% of kinesin-8 deletion cells and once in a wild-type cell (fig. 7C). Although the occurrence of lagging chromosomes was higher in kinesin-8 deletion cells, the difference was not statistically significant given the number of observations ($p = 0.16$ using the Pearson chi-square test for proportions).

The timing of chromosome segregation was significantly altered in kinesin-8 deletion mutants. When we quantified whether sister chromatid segregation had occurred at 5-minute intervals from 0 to 20 minutes after temperature shift (fig. 7D), in wild-type cells, approximately 35% of cells had segregated chromosomes by 20 minutes after temperature shift, while in $5^+6\Delta$ cells, only about 15% of cells had segregated chromosomes by this time and in $5\Delta6^+$ and $5\Delta6\Delta$ cells segregation was never observed. We fit these data to an exponential approach to 1 and found that the segregation time was almost a factor of two larger in $5^+6\Delta$ cells compared to wild type (fig. 7D).

Previous work found alterations in the speed of fission-yeast chromosome segregation in mutants with decreased kinesin-8 localization at KCs (Tang et al., 2014). We measured the distance of each sister chromatid from the SPB to which it was segregating and determined the speed of the final movement toward the SPB (fig. 7E-G). The segregation speeds were similar in wild-type and $5^+6\Delta$ cells, although the small number of observations limited our ability to detect differences.

### 2.7 Kinesin-8 deletion induced aberrant anaphase nuclear envelope deformation

We next asked whether kinesin-8 deletion and associated long mitotic spindles had any effect on nuclear-envelope deformation and anaphase spindle elongation. We used the soluble nuclear GFP visible in wild-type and $5^+6\Delta$ cells from our GFP-lac array-based KC marker to image the shape of the nucleus (fig. 8A-D). Consistent with previous observations in fission yeast (Lim et al., 2007) and our observations of large nuclear-envelope deformation by long mitotic polar MTs in kinesin-8 deletion mutants (fig. 2I), we found that anaphase spindle elongation altered nuclear envelope shape without any apparent breakage of the membrane (fig. 8A-D). However, the nature of nuclear-envelope shape change was sometimes altered in $5^+6\Delta$ cells. For wild-type and some $5^+6\Delta$ cells, the nuclear envelope deformed into a peanut-like shape that was narrowest near the spindle midzone (fig. 8A,C), consistent with previous work (Lim et al., 2007; Lim and Huber, 2009). However, for some $5^+6\Delta$ cells, the nuclear envelope formed a thin tether around the elongating spindle (fig. 8B,C), similar to those observed for long polar MTs (fig. 2I). In previous work, spindle elongation was not associated with tether formation, although other types of nuclear-envelope deformation could produce tethers (Lim et al., 2007; Lim and Huber, 2009).

We quantified speeds of spindle elongation by fitting spindle length versus time (fig. 8E-G). Consistent with previous work (West et al., 2002; Tang et al., 2014), we found similar speeds of spindle elongation in wild-type and kinesin-8 deletion mutant cells.

# 3 Mathematical Model Results

To better understand the KC movements and spindle dynamics observed, we developed a mathematical model of these processes. We wanted to determine the extent to which alterations in catastrophe frequency caused by loss of kinesin-8 proteins could explain the KC pushing movements and spindle length instability we observed. Therefore we developed and tuned a model based on our measurements of KC movements and incorporated those same model parameters into a spindle force balance model.

## 3.1 A model based on microtubule dynamics can reproduce the observed kinetochore movements

We first developed a physical model of an MT bundle that drives movements of an attached KC (fig. 9A, Materials and Methods). We built on previous measurements of the dynamics of single MTs attached to purified budding yeast KCs, which determined the dependence of MT growth speed, shrinking speed, catastrophe frequency, and rescue frequency on the force exerted by an attached KC (Akiyoshi et al., 2010) as well as on previous modeling of MT bundles (Laan et al., 2008) and chromosome oscillations (Gay et al., 2012; Banigan et al., 2015). In our model, dynamic MTs grow and shrink, and MTs can attach to and detach from a KC. For MTs longer than 3 $\mu$m, we added a compressive force to represent the force from nuclear envelope deformation (Lim et al., 2007), and for MTs longer than 10 $\mu$m we added an additional force to represent the rigid cell wall. The KC was initially lost and had no interaction with the bundle until it attaches to the MT bundle. MTs that shortened all the way to the SPB re-nucleated at a variable rate, and we considered various initial MT numbers. We note that we modeled a single bundle: while larger initial MT numbers made the bundle longer and more stable, all MTs moved and searched for a lost KC as a single unit.

We determined model parameters from previous experiments where available and by comparison of the model to our data. To do this, we modeled an imaging experiment and analyzed the simulation data as we did the experiment. This included analysis of speeds and durations of events, bundle length distributions, and the temporal autocorrelation of the movements (Materials and Methods). We varied the model parameters over wide ranges and identified parameter sets that gave the best agreement between the data and the model. This gave us 'cc' parameter sets that represented our wild-type, $5^+6\Delta$, and $5\Delta$ cells (including both $5\Delta6^+$ and $5\Delta6\Delta$). At first, we only allowed the catastrophe frequency to vary between the different parameter sets. However, this difference alone was insufficient to give good agreement between the model and the data. We therefore allowed more parameters to vary until good experiment-model agreement was obtained. Many parameters were the same for all strains (Materials and Methods, table 5). The key differences in parameters corresponding to our different cell types are in catastrophe and rescue frequencies, MT-KC attachment frequency, nucleation frequency, and initial MT number (fig. 9B and table 3).

The wild-type parameters included length-dependent catastrophe, representing the measured increase in catastrophe frequency with MT length in interphase fission-yeast cells due to kinesin-8 motors (Tischer et al., 2009), as well as high catastrophe and rescue frequencies, a low attachment frequency (representing the shorter polar MTs in wild-type cells), and a high nucleation frequency. The average initial number of MTs in the polar bundle was 3, consistent with previous estimates (Ding et al., 1993). With these parameters, the model exhibited rapid KC reeling in to the SPB, relatively short MT-KC lengths, and initially correlated movements for which the correlation decays to zero within 20 minutes (fig. 9C).

Our measurements of KC reeling in for $5^+6\Delta$ cells discussed above were puzzling, because while the lengths of polar MTs were long (as expected for low MT catastrophe frequency), no pushing events were observed and the rates of KC reeling in to the SPB were rapid (as expected for high MT catastrophe frequency). In searching for model parameters representing $5^+6\Delta$ cells, we initially were unable to find parameters that gave both long polar MTs and rapid KC reeling in to the SPB. We hypothesized that this apparent discrepancy could occur due to residual Klp5 activity at KCs, based on previous work that has found that Klp5 can homodimerize (Li and Chang, 2003), a purified Klp5 dimeric construct has a high affinity for tubulin dimers (Erent et al., 2012), and Klp5 localizes to KCs through interactions with Alp7/TACC (Tang et al., 2014). Further, we observed localization of Klp5$^{GFP}$ to spindles in cells lacking Klp6 (fig. 4). Therefore, in the $5^+6\Delta$ parameters, unattached MTs had low catastrophe and rescue frequencies identical to the $5\Delta$ model (discussed below), but attached MTs had more wild-type-like dynamics with increased catastrophe and rescue frequencies (intermediate between wild-type and $5\Delta$, table 3). In addition, consistent with previous work that has found less stable MT-KC attachments in kinesin-8 deletion mutants (Garcia et al., 2002a), we increased the attached MT unbinding rate relative to the wild-type parameters.

When we included these attachment-state-dependent MT dynamics in the model, we found parameters that gave agreement with the $5^+6\Delta$ experimental data if the attachment frequency was high (to model the longer polar MT bundles in $5^+6\Delta$ cells), the nucleation frequency was low (Erent et al., 2012), and the average initial number of MTs in the bundle was 6, consistent with the increased brightness of the KC-associated polar MT bundles in cells of this genotype. While unattached MTs were long, the model exhibited rapid KC reeling in to the SPB, a corresponding short distribution of MT-KC lengths, and initially correlated movements for which the correlation decays to zero within 10 minutes (fig. 9D).

Finally, we determined model parameters representing $5\Delta$ cells which differ from those of the $5^+6\Delta$ model based on the evidence that Klp6 is unable to homodimerize (Garcia et al., 2002b; Li and Chang, 2003) and Klp6$^{GFP}$ shows weak localization to spindles in cells lacking Klp5 (fig. 4). Therefore, we assumed that Klp6 alone was monomeric and unable to promote MT depolymerization. We modeled MTs with low catastrophe and rescue frequencies, independent of the attachment state of the MT (identical to the unattached dynamics in the $5^+6\Delta$ parameters, table 3). Because previous work has found that Klp6 can promote MT nucleation in vitro and in cells lacking Klp5 (Erent et al., 2012), we increased the MT nucleation rate in the $5\Delta$ model. A larger average initial number of MTs in the polar bundle gave the best results. The resulting simulations showed KC pushing movements, larger SPB-KC distances, and correlated movements for which the correlation oscillates and does not decay to zero until 50 minutes (fig. 9E).

## 3.2 A spindle model incorporating force balance and kinesin-8 regulated MT dynamics reproduced metaphase spindle length dynamics

If our MT bundle model accurately described the effects of kinesin-8 deletion of MT dynamics, we expected that adding the parameters we derived for this model into a model of spindle length would reproduce the metaphase spindle length (in)stability that we measured. Therefore, we extended our model to describe spindle length dynamics, using the same MT dynamics and KC attachment/detachment parameters determined above (table 3). Extending the model required that we add the elastic link between sister KCs (Gay et al., 2012; Banigan et al., 2015) and forces on the SPBs due to outward sliding generated by motors on interpolar MTs and nuclear envelope deformation (fig. 10A, Materials and Methods, table 6). We did not explicitly model force generation in the center of the spindle, but included a constant outward sliding force. As for the MT bundle model, we mimicked an imaging experiment and compared characteristics of the data we collected to the simulated experiments.

Initially we attempted to keep all added parameters the same for all cell types (fig. 10B). However, we found that agreement between the model and experiments required that we vary the sliding force that pushes the SPBs apart (table 3). Previous work found destabilization of the spindle interzone in kinesin-8 depleted cells (Gatt et al., 2005), and the budding-yeast kinesin-8 motor Kip3 has antiparallel MT sliding activity that contributes to spindle elongation (Su et al., 2013). Therefore, we used a high sliding force in the wild-type and $5^+6\Delta$ parameter sets (reflecting residual Klp5 homodimer activity) and low force in the $5\Delta$ parameter set.

With this one additional parameter that varies, we found that the model remarkably reproduced spindle length fluctuations. The wild-type parameters led to typical spindle lengths of 2-3 $\mu$m that were stable over time (fig. 10C) as well as increased probability for KCs to be located near the center of the spindle (fig. 10D). The $5^+6\Delta$ parameters gave intermediate spindle stability: in some cases spindle length was relatively stable, while in others the spindle length showed large fluctuations (fig. 10E). Consistent with our experimental results, the $5^+6\Delta$ parameters led to decreased KC centering on the spindle (fig. 10F). The $5\Delta$ parameters produced consistently large fluctuations in spindle length, similar to those we observed experimentally (fig. 10G) and a loss of KC centering (fig. 10H).

## 4 Discussion

Despite recent work on the effects of kinesin-8 motors on MT dynamics, the connection between alterations in MT dynamics and kinesin-8 mitotic phenotypes is not clear, in part because of the many reported effects of kinesin 8s on MTs and mitosis (Erent et al., 2012; Messin and Millar, 2014). Therefore, we studied effects of kinesin-8 deletion on chromosome movements and spindle length dynamics in fission yeast. We wanted to know the extent to which kinesin-8 mitotic phenotypes can be understood based on the known effects of these motors on MT catastrophe, or if other kinesin-8 activities are important. As discussed in more detail below, comparison of our experimental data to results of mathematical modeling suggest that the length-dependent increase in MT catastrophe frequency of Klp5/6 (Tischer et al., 2009) is important both for normal chromosome movements and for spindle length stability. In addition, our best model of wild-type cells incorporates Klp5/6dependent contributions to (1) the MT rescue frequency (Unsworth et al., 2008), which helps stabilize MTs in polar bundles and in the spindle; (2) stabilization of MT-KC attachments (Garcia et al., 2002a), which helps maintain the mechanical connections between MTs and KCs that reel lost KCs in to the SPBs and stabilize the spindle; (3) the MT nucleation frequency (Erent et al., 2012), which helps regulate the number of MTs in polar MT bundles; and (4) the sliding force that elongates the spindle (Su et al., 2013), which is important for spindle length regulation.

In fission-yeast cells with induced lost KCs, we observed reattachment of KCs to MTs, reeling in of KCs to the SPBs, biorientation, and segregation (fig. 1, 2). Cells lacking Klp5 and Klp6 exhibited aberrant chromosome movements: KCs sometimes hovered near the reeling SPB, and cells lacking Klp5 showed MT-driven pushing of KCs away from the SPB. The time for the KC to reel in to the SPB was shorter in 5$^+$6Δ cells than in wild type but longer in 5Δ cells. Reeling speeds were also slower in 5Δ cells. These differences cannot be understood solely through alterations in the MT catastrophe frequency upon kinesin-8 deletion, because the different deletions had different effects.

In addition to the aberrant chromosome pushing movements, occasional cells lacking Klp5 contained three SPBs and tripolar mitotic spindles (fig. 3). Typically one of the three SPBs appeared dimmer than the other two. We observed what appeared to be formation of a tripolar spindle by splitting of one SPB into two (fig. 3C), which suggested the possibility that the highly stable MTs in 5Δ6$^+$ cells could exert forces large enough to split an SPB. However, cells with aberrant KC movements were more likely to have total KC intensity much brighter or dimmer than average (fig. 3D), suggesting that variation in the chromosome and SPB number could be associated with aberrant dynamics.

Our results were surprising given that previous work showed similar mitotic phenotypes for deletion of either Klp5 or 6 (West et al., 2002; Unsworth et al., 2008) and loss of nuclear localization of Klp5/6 when the partner motor was absent (Unsworth et al., 2008), although other work did not see this (West et al., 2002). We re-examined Klp5/6 localization in the presence and absence of the partner motor (fig. 4). When Klp5 was present without Klp6, it localized to the spindle and polar MT ends. When Klp6 was present without Klp5, it showed weaker spindle and polar MT localization but consistent localization near SPBs. These results suggest that either Klp5 or 6 can still be present in the nucleus at low levels in the absence of the partner protein. The stronger localization of Klp5 alone versus Klp6 alone is consistent with previous work that found that Klp5 can form homodimers but Klp6 cannot (Garcia et al., 2002b; Li and Chang, 2003).

After KC reeling in to the SPB, in 5$^+$6Δ cells biorientation occurred at a normal time, but in 5Δ cells, biorientation occurred much more slowly, if at all (fig. 5). In 5$^+$6Δ cells, the separation of sister KCs was reduced relative to wild type, suggesting a decrease in KC tension in agreement with most other measurements (Garcia et al., 2002a; Zhu and Jiang, 2005; Mayr et al., 2007; Stumpff et al., 2008; Wargacki et al., 2010) but contradicting previous work on fission yeast (West et al., 2002). After biorientation in 5$^+$6Δ cells, the marked chromosome stayed closer to the pole that reeled it in, suggesting an impairment of either elongation of the reeling MT, attachment to MTs from the other SPB, or shortening of MTs from the other SPB.

Deletion of either kinesin 8 led to instability in spindle length, suggesting that this mutation caused a weakening or loss of spindle length stabilization (fig. 6). The effect was more marked in 5Δ cells, consistent with the more severe phenotypes observed upon deletion of Klp5. Speeds of elongation and shortening were similar to wild type, but the amplitude of the length fluctuations was much greater, and the average spindle length was considerably longer. Spindle length decreases were previously reported in fission-yeast cells with kinesin-8 or Dam1 deletions (Syrovatkina et al., 2013) or a mutation in the Ndc80 loop which is defective in Dis1/XMAP215 recruitment to KCs (Hsu and Toda, 2011). Because XMAP215 proteins typically promote MT polymerization, this suggests that perturbations to MT dynamics can destabilize spindle length, but that this cannot be understood by a simple effect on MT length or catastrophe frequency. Recent work has examined the mechanisms responsible for the control of average spindle length in a range of organisms (Dumont and Mitchison, 2009; Goshima and Scholey, 2010; Wühr et al., 2008; Hara and Kimura, 2009; Civelekoglu-Scholey et al., 2010; Greenan et al., 2010; Good et al., 2013; Hazel et al., 2013; Hara and Kimura, 2013; Reber et al., 2013; Su et al., 2013; Syrovatkina et al., 2013; Hepperla et al., 2014; Nannas et al., 2014; Rizk et al., 2014). However, little is known about the mechanisms associated with stabilization of spindle length over time.

The cells with kinesin-8 deletion were delayed in starting anaphase, consistent with previous measurements (West et al., 2002; Garcia et al., 2002b,a; Mayr et al., 2007) and observations that the spindle-assembly checkpoint remains active under kinesin-8 deletion or depletion (Garcia et al., 2002a; Mayr et al., 2007; Meadows et al., 2011). The effect is strongest in 5Δ cells (fig. 7). Klp5/6 binds protein phosphatase 1 at KCs, and while Klp6 has two PP1 binding motifs, while Klp5 has only one (Tang et al., 2014). Since PP1 recruitment to KCs affects the timing of mitotic progression, the longer mitotic delays we measured in 5Δ6+ cells could be explained by differences in PP1 binding. In addition, the elongated K-fibers found in these strains may generate insufficient inter-KC tension due to not pulling on the KCs, a phenomenon consistent with kinesin 8s promoting MT depolymerization and Klp5 being more important for this function. In anaphase, kinesin-8 deletion increased the fraction of lagging chromosomes, and again Klp5 deletion had the bigger effect. This observation is consistent with a K-fiber that cannot shorten properly due to impaired depolymerization. In addition, we found a surprising phenotype that in kinesin-8 deletion mutants, the anaphase nuclear-envelope deformation in kinesin-8 deletion mutants sometimes changed from the typical peanut-like shape to formation of a tether around the anaphase spindle (fig. 8).

We found that deletion of Klp5 and 6 can result in different mitotic phenotypes. Most previous work has not observed differences during fission-yeast vegetative growth. However, it has long been known that Klp5 and 6 show significantly different meiotic phenotypes: the spore viability of zygotic asci differed significantly for crosses between 5/6 single/double deletes and wild-type cells, and Klp5/6 deletion homozygous crosses produced different number and morphology of spores (West et al., 2001). Because Klp5 and 6 have been understood to function together as a heterodimer, much previous mitotic work has looked at effects of perturbing only Klp5 or 6. However, some previous work has hinted at differences between Klp5 and 6 mitotic function. A study of the Cdc14-related phosphatase Clp1 found that both Klp5 and 6 are detectable by mass spectrometry after affinity capture by Clp1, but only Klp6 is detectable by Western blot after pulldown. Additionally, only Klp6 is able to reciprocally affinity capture and be dephosphorylated by Clp1 (Chen et al., 2013).

Similarly, the KASH protein Kms2 interacted with Klp5 but not 6 by mass spectrometry after affinity capture (W¨alde and King, 2014). A recent high-throughput study quantified the effects of paired gene deletions in fission yeast on colony growth (Ryan et al., 2012). This study found 298 genes with a significant interaction score (magnitude>1) for Klp5/6 but not the partner, indicating that many differences in genetic interactions for Klp5 and 6 occur. Mitotic and microtubule-related genes that showed significant differences in genetic interaction with Klp5 and Klp6 included Ase1, EB1/Mal3, CLIP-170/Tip1, Tea2, and the gamma-tubulin complex subunit Alp16.

We developed a mathematical model of MT-driven KC movements (fig. 9) and metaphase spindle length fluctuations (fig. 10) that gives a consistent interpretation of our results based on a conceptual model of Klp5/6 activity (fig. 11). In this picture, Klp5/6 heterodimers form in wildtype cells and have strong binding and activity on MTs and at KCs. The catastrophe frequency is high and length dependent (Tischer et al., 2009), the rescue frequency is high (Unsworth et al., 2008), MT-KC attachments are stable (Garcia et al., 2002a), the sliding force that elongates the spindle is high (Su et al., 2013), the MT nucleation frequency is high (Erent et al., 2012), and polar MT bundles typically contain 3 MTs. As a result, polar MTs rapidly reel in lost KCs to SPBs, spindles and polar MTs are at their wild-type regulated lengths, spindle length is stable in time, and KCs center on the spindle (fig. 11A).

If Klp5 can homodimerize (Garcia et al., 2002b; Li and Chang, 2003), it may reach MT plus ends either by motor activity or by binding tubulin dimers (Erent et al., 2012), where it might retain some KC localization and activity through interactions with Alp7/TACC at the KC (Tang et al., 2014). Residual Klp5 activity would explain why 5$^+$6Δ cells show less severe phenotypes and stronger spindle localization in our experiments. Perhaps the Klp5 homodimer activity is unregulated, making depolymerization faster during reeling in to the SPB. We propose that in 5$^+$6Δ cells active Klp5 homodimers are present at KCs, leading KC-attached MTs to have more wild-type dynamics with higher catastrophe and rescue frequencies, while unattached MTs have low catastrophe and rescue frequencies. This interpretation may explain why polar MTs in 5$^+$6Δ cells are long (fig. 2L) but SPB-KC distances during reeling in are short (fig. 9D). In our mathematical model the 5$^+$6Δ parameter set leads to both rapid reeling in of lost KCs to SPBs but also longer spindles and polar MTs, variable spindle length stability (sometimes stable, sometimes fluctuating), and loss of KC centering on the spindle (fig. 11B).

If Klp6 cannot homodimerize (Garcia et al., 2002b; Li and Chang, 2003), its spindle localization may be due to monomer binding, but it might still bundle MTs and promote their nucleation due to its tail MT binding site (Erent et al., 2012). This interpretation may explain why 5Δ cells show more severe phenotypes and weaker spindle localization in our experiments. Therefore, our 5Δ mathematical model assumes no effect of Klp6 on MT dynamics either at KCs or on other MTs, and correspondingly the sliding force that elongates the spindle is decreased, consistent with loss of kinesin-8 sliding activity (Su et al., 2013). The 5Δ parameters also reflect a high MT nucleation frequency either due to direct effects of monomeric Klp6 on nucleation (Erent et al., 2012) or to KCs stuck at SPBs (Garcia et al., 2002b). These parameters lead to both reeling movements and pushing of KCs away from SPBs, long spindles and polar MTs, unstable spindle length, and loss of KC centering on the spindle (fig. 11C).

We note that the decrease in the sliding force parameter is necessary for the 5Δ spindle model to agree with our experimental results, assuming that other parameters are the same in the bundle and spindle models. While this is consistent with Klp5/6 possessing a sliding activity, there could be other explanations. For example, Klp5/6 might contribute to outward sliding force generation in the spindle indirectly, or the sliding force might unchanged but multiple other parameters change instead (although that we were not able to find such parameters).

While differences in homodimerization of Klp5 and 6 offer one possible explanation for our results, our data do not rule out other interpretations. Klp5 and 6 may interact with different proteins, as has been observed for Clp1 (Chen et al., 2013) and Kms2 (W¨alde and King, 2014). These differences in interactions might explain the different phenotypes we observe when deleting Klp5 or 6.

Previous mathematical models of spindles in yeasts have assumed fixed spindle length (Sprague et al., 2003; Gardner et al., 2005; Pearson et al., 2006; Gardner et al., 2008; Gay et al., 2012; Hepperla et al., 2014). To our knowledge, our work is the first mathematical model to display stability or instability of spindle length depending on parameters.

The parameters determined for our mathematical model developed to explain KC movements (fig. 9) were used directly to model metaphase spindle length fluctuations (fig. 10). Remarkably, the model parameters we determined based on KC movements led directly to the differences in spindle length stability we observed experimentally, if we also postulated a kinesin-8 dependent sliding force. This suggests that alterations in catastrophe, rescue, and nucleation frequency, MTKC attachment stability, and sliding force capture the essential cellular roles of Klp5/6 required to explain our observations.

# 5 Materials and Methods

## 5.1 Experimental methods

### 5.1.1 Strain construction and preparation for microscopy

Cells were cultured using standard techniques (Moreno et al., 1991). For strain construction, existing strains were crossed and the desired phenotype isolated using random spore analysis. Colonies were screened for motor deletions using PCR with primers unique to each deletion (West et al., 2001; Grishchuk and McIntosh, 2006). The *nda3-KM311* (Hiraoka et al., 1984; Kanbe et al., 1990) mutation was identified by replica plating colonies onto YE5S plus phloxin B agar plates. After 3 days at 18°C, dark pink colonies were identified as positive for nda3. The cen2-GFP marker of the centromere on chromosome 2 (Yamamoto and Hiraoka, 2003), the pcp1-GFP spindle pole body marker (Flory et al., 2002), and mCherry-atb2 microtubule marker (Yamagishi et al., 2012) were identified using fluorescence microscopy. Because the *klp5* deletion and the cen2-GFP tag were both marked with $ura4^+$, the identification of these strains was done by imaging and PCR, not by marker selection.

Once constructed, strains were grown at 32°C on YE5S agar plates plus 0.05 M thiamine (Sigma-Aldrich, St. Louis, MO) to reduce the bright signal of the pcp1-GFP marker by partially suppressing the nmt1 promoter (Maundrell, 1990). After a large number of cells reached exponential growth, the plates were moved to a temperature-controlled incubator and kept between 17.5 and 18.5°C. The cells remained at restrictive temperature for 7-9 hours to allow cells to accumulate at M-phase but without MTs and to permit the KCs to separate from the SPBs and move within the nucleus. Cells were then prepared for microscopy. Cells with kinesin-8 deletions are susceptible to chromosome missegregation and other mitotic abnormalities. To avoid problems resulting from these abnormalities, cells for each experiment were re-isolated from frozen stocks on a weekly basis.

### 5.1.2 Viability experiment

Strains were screened for viability after exposure to restrictive temperatures on both liquid and solid media by exposing them to 18°C for 9 hours, returning them to room temperature and observing growth after recovery. Rates of growth were observed qualitatively on agar plates and quantitatively in liquid media by measuring the optical density at 560 nm at 6 time points (table 1).

### 5.1.3 Widefield live-cell imaging

A 22x40 mm coverslip was glow discharged and coated with 4 *µ*L of lectins from *Bandeiraea simplicifolia* (Sigma-Aldrich, St. Louis, MO). 200 *µ*L of YE5S medium was placed onto the coverslip and cooled to 18°C. A small volume of exponentially growing cells was removed from the petri dish and placed on the coverslip. The cells were allowed to settle for 10 minutes to adhere to the lectins. The coverslip was then washed twice with precooled medium, inverted and mounted onto slides with double-sided tape acting as a spacer. The chamber was filled with cooled media and moved to the microscope. An Axioplan II light microscope (Carl Zeiss, Jena, Germany) was equipped with a 100x, 1.45 NA Plan Fluor oil-immersion objective and a Photometrics Cascade 650 CCD camera (Roper Scientific, Sarasota, FL) and was used in a room cooled to 18°C. The objective was equipped with two heating options: a custom water jacket connected to a 32°C water bath for quick heating, and an electric objective heater capable of measuring and maintaining a warm objective temperature (Bioptechs, Butler, PA). Samples were screened with 3-5% lamp intensity at 488nm to identify cells in which the KC of chromosome 2 had separated from the spindle pole bodies. Once a cell was identified, the heating water jacket and objective heater were turned on. After 90 sec, when the permissive temperature of 32ºC was reached in the slide (determined based on calibration with a slide-mounted thermistor), the heating water jacket was turned off and the temperature was maintained by the objective heater. We collected image stacks at 15 sec intervals (75 ms exposure, 11-13 focal planes, 200 nm step size between focal planes) until the KC was reeled in to the SPB, the signal was photobleached, or the region of interest was lost. The stacks were deconvolved with the nearest neighbor function in Metamorph 6.3 Image Analysis Software (Molecular Devices, Sunnyvale, California), and each z-stack was displayed as a maximum intensity projection.

### 5.1.4 Confocal imaging

Cells were prepared at 18ºC as above. For fixed-cell preparations, cells were fixed with methanol at -20ºC for 8 minutes and washed 3 times with PEM buffer (100 mMol PIPES, 1 mMol EGTA, 1 mMol MgSO4). The cells were mounted onto lectin-coated coverslips and stained with DAPI. For live-cell preparations, cells were transferred into 35 mm glass-bottom dishes (MatTek, Ashland, MA) containing 200 $\mu$L of YE5S media and placed onto a spinning disk (Yokogawa, Musashino, Japan) TE2000U confocal inverted microscope (Nikon, Tokyo, Japan) with a 100X, 1.4 NA Plan Apo oil-immersion objective and an ORCA EM CCD camera (Hamamatsu, Hamamatsu, Japan). The stage was pre-warmed to 32ºC, and images were taken with 750-1000 ms exposure for mCherry and 100 ms exposure for GFP. Additional images were collected on a spinning disk (Yokogawa, Musashino, Japan) Cell Voyager CV1000 confocal inverted microscope (Olympus, Tokyo, Japan) with a 100X, 1.4 NA Plan Apo oil-immersion objective and a C9100-23B EM-CCD camera (Hamamatsu, Hamamatsu, Japan). Live-cell images of the Klp5GFP and Klp6GFP tagged strains were taken on a spinning disk (Yokogawa, Musashino, Japan) Ti Microscope (Nikon, Tokyo, Japan) with a 100x, 1.45 NA Plan Apo oil-immersion objective and a C9100-13 EM-CCD camera (Hamamatsu, Hamamatsu, Japan). Other live-cell images (Fig. 3, Fig 4C-F) were taken on the same microscope but with an Andor iXon Ultra 897 EM-CCD camera (Andor, Belfast, United Kingdom). Images from all confocal microscopes were displayed as pixel interpolated maximum-intensity projections.

### 5.1.5 Spot tracking and analysis

Images and movies shown were background subtracted and filtered to allow for signal tracking using the Fiji version of Image J (NIH, Bethesda, MD). Spots in the processed stacks were tracked in three dimensions using the TrackMate v2.2.0 plugin in Fiji. The spindle poles and the KC were then tracked to generate time series of 3D positions. The data were analyzed using custom Matlab routines (MathWorks, Natick, MA, version 8.5).

Phenotype classification was performed by selecting 20-30 cells of each genotype (numbers in fig. 2M) for which a lost KC could be identified at the time of temperature shift and imaging occurred for at least 15 minutes. In each cell of the population we recorded the occurrence of phenotypes, including pushing and hovering (fig. 2J) and lagging chromosome movement at anaphase (fig. 7C). Additionally at 0, 5, 10, 15, and 20 minutes after temperature shift each cell was scored for whether or not KC reeling to the SPB (fig. 2K), biorientation (fig. 5C), and chromosome segregation (fig. 7D) had occurred.

Analysis of speeds of reeling movements (fig. 2N,O), chromosome segregation (fig. 7E–G), and anaphase spindle elongation (fig. 8E–G) were done by hand-selecting start and end points of events, then performing a least-squares linear fit to the length versus time during the event.

To Automatically identify spindle length (table 2) and SPB-KC distance (fig. 9C–E) change events, we first by performed a running average of 5 timepoints to smooth the data, identified successive local maxima and minima of the smoothed curve, and defined events as intervals between successive local maxima and minima of at least 105 seconds (7 time points). Length changes and durations of each event were determined from the smoothed curve. The raw length (or distance) versus time points in each event were fit to a line to determine speeds.

The temporal autocorrelation of fluctuations in spindle length (fig. 6D) and SPB-KC distance (fig. 9C–E) were calculated for each experiment by computing the normalized autocorrelation function and its variance using the autocorr function in Matlab, then dividing by the square of the mean length (or distance) of the experiment. The single experiment autocorrelation functions were averaged weighted by the number of points for each time delay for each experiment.

### 5.1.6 Brightness analysis of 5Δ strains

Crosses creating 5Δ strains often contained multiple copies of one or more chromosomes, so strains and cells were chosen for imaging that lacked multiple cen2-GFP dots. Additionally, 5Δ strains exhibited decreased cen2-GFP signal brightness compared to wild-type and $5^+6\Delta$, apparently because of variation from strain to strain in the soluble pool of GFP. This dimness made it difficult to identify the cen2-GFP compared to other dim GFP signals, which appeared to be aggregrates of GFP, as has been observed for this construct in other fission-yeast mutants (Ayumu Yamamoto, personal communication). Numerous strains lacking klp5 and containing cen2-GFP were constructed, all with a similar brightness phenotype. This decrease in the soluble pool of GFP available to bind to the cen2 was attributed to the the genetic linkage between klp5, cen2, and his7 (Yamamoto and Hiraoka, 2003). To avoid possible imaging artifacts associated with soluble GFP aggregates, we performed an analysis of all 5Δ cells with lost KCs. We grouped cells into categories called interacting (the putative cen2 showed signs of MT-driven interactions with an SPB), lost (the putative cen2 never clearly interacted with an SPB), and junk (the putative cen2 was very dim or bleached quickly). The brightness of all dots was tracked and measured over time using the Manual Tracking plugin in Fiji, and histograms of the ratio of dot intensity to SPB intensity were compared. This analysis showed a clear difference in brightness between interacting and junk dots, with junk dots being much dimmer. Lost dots showed a bimodal intensity ratio distribution, with peaks that matched those of the interacting and junk dots. We therefore set a cut-off at the minimum of the histogram between the two peaks, and threw out all dots with dot/SPB intensity below the cutoff.

### 5.1.7 Polar MT length measurements

The lengths of polar MTs were measured from maximum intensity projections of 3D time-lapse images from the Cell Voyager CV1000 microscope described above for wild-type, $5^+6\Delta$ and $5\Delta6^+$ cells taken at 1 minute intervals from 0 minutes to 10 minutes after exposure to 32°C. Straight lines were drawn from the center of the SPB to the visible end of the brightness/contrast optimized MT bundle and the length was obtained using the measure function in Image J.

### 5.1.8 Determination of $5\Delta6^+$ cell chromosome copy number

To determine whether $5\Delta6^+$ cells were stably diploid, we first streaked the cells into single colonies and replica plated onto YE5S plus phloxin B agar plates. After 24, 48 and 72 hours of growth, colonies were a uniform light pink color (rather than the dark red typical of diploid strains), indicating they were haploid. We also attempted to induce sporulation of $5\Delta6^+$ strains by streaking cells onto malt extract agar plates and attempting to visualize azygotic asci after 72 hours. None were seen, indicating the strains were haploid.

To test whether individual cells might have altered numbers of KCs and SPBs, we collected time-lapse images of 5Δ6+ cells, including those of fig. 3. Movies were divided into groups that showed apparently normal KC dynamics and aberrant dynamics. We then quantified the total KC brightness in each frame by first performing 3D point source detection and fitting 3D Gaussians to each spot in our image (Thomann et al., 2002). We then used the u-Track tracking algorithm with Kalman filtering, gap closing, splitting, and merging to link features between frames (Jaqaman et al., 2008), where we required a minimum of two linked frames to identify persistent fluorescent spots. The total KC intensity for each frame was determined by summing the 3D Gaussian amplitudes for all persistent fluorescent spots.

## 5.2 Mathematical models

### 5.2.1 Microtubule bundle model

The 1D MT bundle model builds on previous work modeling dynamic instability of MT bundles (Laan et al., 2008), extended to include a KC elastically linked to attached MTs (Gay et al., 2012; Banigan et al., 2015), leading to force-dependent MT dynamics (Akiyoshi et al., 2010). In our model, MTs undergo dynamic instability with a growth speed, catastrophe frequency, shrinking speed, and rescue frequency that are force dependent. Following the results of Dogterom and Yurke, the MT growth speed has the form $v_g = v_1 e^{F/F_{cg}} - v_2$ for compressive forces, where $F$ is the force applied to the MT tip along its axis in the direction of the MT plus end (positive for tension, negative for compression) (Dogterom and Yurke, 1997). The constants $v_1$ and $v_2$ are chosen to recover the growth speed at zero force and a growth speed of zero at the stall force. For MTs under tension the growth speed remains the zero-force growth speed to avoid numerical inaccuracy due to exponential growth of the speed with force. The shrinking speed is $v_s = v_{s0} e^{-F/F_{cs}}$, the catastrophe frequency is $f_c = f_{c0} e^{-F/F_{cc}}$, and the rescue frequency is $f_r = f_{r0} e^{F/F_{cr}}$, as measured by Akiyoshi et al. (2010). Any MT which grows to within an interaction distance of the KC becomes attached, and bound MTs unbind from the KC at force-dependent rates for growing ($k_{ug} = k_{ug0} e^{F/F_{cu}}$) and shrinking ($k_{us} = k_{us0} e^{-F/F_{cu}}$) states. MTs that shrink to zero length renucleate in the growing state at the nucleation frequency.

When attached, the KC experiences a linear elastic force from all MTs with which it interacts. When the SPB-KC length is greater than 3 $\mu$m (the nuclear diameter), it experiences a constant poleward force representing the force to deform the nuclear envelope (Lim et al., 2007; Lim and Huber, 2009). When the SPB-KC length is greater than 10 $\mu$m, the poleward force increases by a factor of 10 to represent the force of hitting the cell wall. When unattached, the KC undergoes a Gaussian-distributed random displacement at each timestep with standard deviation determined from the KC diffusion coefficient (Kalinina et al., 2012). At the start of each simulation, the KC begins in an noninteracting state in which it is unable to interact with MTs; it switches to the interacting state at the attachment rate. In addition to the linear elastic force experienced by attached MTs, any unattached MTs that grow past the KC experience an additional linear elastic force representing interactions with the chromosome that oppose MT growth.

We simulated the model using a kinetic Monte Carlo algorithm in which, at each time step, MT length is updated according to the growing/shrinking speeds for each MT. MTs change state with the appropriate catastrophe/rescue and binding/unbinding frequencies, and forces on MTs are updated. Because the drag force on the KC is small compared to other forces in the problem, the KC moves to the instantaneous position of zero force at each time step. The initial MT length and SPB-KC distance were both set to 1 $\mu$m. MTs were initially in the growing state and unattached from the KC. Each simulation was run for 60 minutes with a time step of $2 \times 10^{-3}$ min. If the KC was reeled in to the SPB and remained there for more than 3 minutes, the simulation was ended. For comparison to experimental distributions, 30 simulation traces were included in an experiment. Simulated length versus time traces were made by taking the simulation results, sampling every 15 seconds to match the experimental imaging time interval, and adding Gaussian-distributed random noise to each point with a standard deviation of approximately 125 nm.

To compare the experimental and model results, we performed sweeps of parameter space (varying combinations of 4-9 parameters over 1-2 orders of magnitude). The key parameters that were varied were the four dynamic instability parameters (growth speed, shrinking speed, catastrophe frequency, and rescue frequency), the number of MTs in a bundle, the MT polymerization stall force, the unbinding rate of the kinetochore-MT attachment from growing and shrinking MTs, and the nucleation rate. We defined an error function that compared key features of the model to the data, and studied how the model output and the error function changed with parameters. The error function included analysis of speeds and durations of events, bundle length distributions, and the temporal autocorrelation of the movements. In addition to the parameter sweeps, we performed global optimization using matlab's pattern search algorithm to search for best-fit parameters that minimized the error function. The parameters we presented in the paper were the best fits for the given datasets to which we compared, as determined both by optimization and parameter sweeps.

### 5.2.2 Spindle force balance model

The spindle force balance model is related to previous work modeling force balance in the spindle (Cytrynbaum et al., 2003, 2005). We extended the MT bundle model with the following additions and changes. Multiple sister KC pairs are present and a fixed number of MTs can attach to each sister KC. Sister KCs are linked by a linear elastic spring. If pushing forces from MTs cause sister KCs to approach closer than the chromosome offset distance, an additional linear elastic force opposing KC overlap is added. MTs that grow into the opposite SPB experience a linear force opposing the MT-SPB overlap. To maintain a fixed number of spindle MTs, any MTs that shrink to zero length are renucleated in the growing state. The maximum MT shrinking speed is set to 30 $\mu$m/min to avoid numerical instability in the simulations.

SPBs experience a constant pushing force directed away from the center of the nucleus that represents the force exerted by sliding motors on interpolar MTs. When the SPBs are separated by a distance greater than the diameter of the nucleus, they experience a centering force representing the force to deform the nuclear envelope of the form $F_{NE} = F_{max}(1-e^{-|s_{spb}-r_{nucl}|/0.2\mu m})$, chosen to smoothly increase to the maximum value.

In the kinetic Monte Carlo algorithm, at each time step all KCs and both SPBs are moved to the instantaneous position of zero force on each object. If an SPB loses all MT attachments to KCs, it moves under the external forces it experiences based on the SPB drag coefficient. The initial MT length was 100 nm, spindle length was 200 nm, and KCs were centered on the spindle. MTs were initially in the growing state and unattached from the KCs. Each simulation was run for 30 minutes with a time step of $5\times10^{-3}$ min. For comparison to experimental distributions, simulated length versus time traces were made by sampling the simulation results and adding noise, as for the pushing bundle model.

**Acknowledgements**. We thank Julie Cooper, Jonathan Millar, and Takashi Toda for providing fission yeast strains, and Chip Asbury, Vanja Dukic, Greg Huber, Andrea Liu, and Shelley Sazer for useful discussions. This work was supported by NSF grant DMR-0847685 (MB), NIH grant K25 GM110486 (MB) and NIH grants R01 GM033787 (JRM). This worked used code from the Danuser and Jaqaman labs, provided through the Computational Image Analysis in Cellular and Developmental Biology course at MBL, funded by NIH grant R25 GM103792.## References

Akiyoshi, B., Sarangapani, K. K., Powers, A. F., Nelson, C. R., Reichow, S. L., Arellano-Santoyo, H., Gonen, T., Ranish, J. A., Asbury, C. L., and Biggins, S. (2010). Tension directly stabilizes reconstituted kinetochore-microtubule attachments. *Nature*, 468(7323):576–579.

Banigan, E. J., Chiou, K. K., Ballister, E. R., Mayo, A. M., Lampson, M. A., and Liu, A. J. (2015). Minimal model for collective kinetochore–microtubule dynamics. *Proceedings of the National Academy of Sciences*, 112(41):12699–12704.

Chen, J.-S., Broadus, M. R., McLean, J. R., Feoktistova, A., Ren, L., and Gould, K. L. (2013). Comprehensive proteomics analysis reveals new substrates and regulators of the fission yeast Clp1/Cdc14 phosphatase. *Molecular & Cellular Proteomics*, page mcp.M112.025924.

Civelekoglu-Scholey, G., Tao, L., Brust-Mascher, I., Wollman, R., and Scholey, J. M. (2010). Prometaphase spindle maintenance by an antagonistic motor-dependent force balance made robust by a disassembling lamin-B envelope. *The Journal of Cell Biology*, 188(1):49–68.

Cottingham, F. R. and Hoyt, M. A. (1997). Mitotic spindle positioning in Saccharomyces cerevisiae is accomplished by antagonistically acting microtubule motor proteins. *The Journal of Cell Biology*, 138(5):1041–1053.

Cytrynbaum, E., Scholey, J., and Mogilner, A. (2003). A Force Balance Model of Early Spindle Pole Separation in Drosophila Embryos. *Biophysical Journal*, 84(2):757–769.

Cytrynbaum, E. N., Sommi, P., Brust-Mascher, I., Scholey, J. M., and Mogilner, A. (2005). Early Spindle Assembly in Drosophila Embryos: Role of a Force Balance Involving Cytoskeletal Dynamics and Nuclear Mechanics. *Molecular Biology of the Cell*, 16(10):4967–4981.

De Wever, V., Nasa, I., Chamousset, D., Lloyd, D., Nimick, M., Xu, H., Trinkle-Mulcahy, L., and Moorhead, G. B. G. (2014). The human mitotic kinesin KIF18A binds protein phosphatase 1 (PP1) through a highly conserved docking motif. *Biochemical and Biophysical Research Communications*, 453(3):432–437.

DeZwaan, T. M., Ellingson, E., Pellman, D., and Roof, D. M. (1997). Kinesin-related KIP3 of Saccharomyces cerevisiae is required for a distinct step in nuclear migration. *The Journal of Cell Biology*, 138(5):1023–1040.

Ding, R., McDonald, K. L., and McIntosh, J. R. (1993). Three-dimensional reconstruction and analysis of mitotic spindles from the yeast, Schizosaccharomyces pombe. *The Journal of Cell Biology*, 120(1):141–151.

Dogterom, M. and Yurke, B. (1997). Measurement of the Force-Velocity Relation for Growing Microtubules. *Science*, 278(5339):856–860.

Du, Y., English, C. A., and Ohi, R. (2010). The kinesin-8 Kif18A dampens microtubule plus-end dynamics. *Current Biology*, 20(4):374–380.22

Jaqaman, K., Loerke, D., Mettlen, M., Kuwata, H., Grinstein, S., Schmid, S. L., and Danuser, G. (2008). Robust single-particle tracking in live-cell time-lapse sequences. *Nature Methods*, 5(8):695–702.

Kalinina, I., Nandi, A., Delivani, P., Chac´on, M. R., Klemm, A. H., Ramunno-Johnson, D., Krull, A., Lindner, B., Pavin, N., and Toli´c-Norrelykke, I. M. (2012). Pivoting of microtubules around the spindle pole accelerates kinetochore capture. *Nature Cell Biology*.

Kanbe, T., Hiraoka, Y., Tanaka, K., and Yanagida, M. (1990). The transition of cells of the fission yeast beta-tubulin mutant nda3-311 as seen by freeze-substitution electron microscopy. Requirement of functional tubulin for spindle pole body duplication. *Journal of Cell Science*, 96(2):275–282.

Kim, H., Fonseca, C., and Stumpff, J. (2014). A unique kinesin-8 surface loop provides specificity for chromosome alignment. *Molecular Biology of the Cell*, 25(21):3319–3329.

Kuan, H.-S. and Betterton, M. D. (2013). Biophysics of filament length regulation by molecular motors. *Physical Biology*, 10(3):036004.

Laan, L., Husson, J., Munteanu, E. L., Kerssemakers, J. W. J., and Dogterom, M. (2008). Forcegeneration and dynamic instability of microtubule bundles. *Proceedings of the National Academy of Sciences*, 105(26):8920–8925.

Li, Y. and Chang, E. C. (2003). Schizosaccharomyces pombe Ras1 effector, Scd1, interacts with Klp5 and Klp6 kinesins to mediate cytokinesis. *Genetics*, 165(2):477–488.

Lim, G. H. W. and Huber, G. (2009). The Tethered Infinitesimal Tori and Spheres Algorithm: A Versatile Calculator for Axisymmetric Problems in Equilibrium Membrane Mechanics. *Biophysical Journal*, 96(6):2064–2081.

Lim, G. H. W., Huber, G., Torii, Y., Hirata, A., Miller, J., and Sazer, S. (2007). Vesicle-Like Biomechanics Governs Important Aspects of Nuclear Geometry in Fission Yeast. *PLoS ONE*, 2(9):e948.

Masuda, N., Shimodaira, T., Shiu, S.-J., Tokai-Nishizumi, N., Yamamoto, T., and Ohsugi, M. (2011). Microtubule Stabilization Triggers the Plus-End Accumulation of Kif18A/kinesin-8. *Cell Structure and Function*, 36(2):261–267.

Maundrell, K. (1990). nmt1 of fission yeast. A highly transcribed gene completely repressed by thiamine. *Journal of Biological Chemistry*, 265(19):10857–10864.

Mayr, M. I., Hu¨mmer, S., Bormann, J., Gru¨ner, T., Adio, S., Woehlke, G., and Mayer, T. U. (2007). The Human Kinesin Kif18A Is a Motile Microtubule Depolymerase Essential for Chromosome Congression. *Current Biology*, 17(6):488–498.

Mayr, M. I., Storch, M., Howard, J., and Mayer, T. U. (2011). A Non-Motor Microtubule Binding Site Is Essential for the High Processivity and Mitotic Function of Kinesin-8 Kif18A. *PLoS One*, 6(11):e27471.

Meadows, J. C., Shepperd, L. A., Vanoosthuyse, V., Lancaster, T. C., Sochaj, A. M., Buttrick, G. J., Hardwick, K. G., and Millar, J. B. (2011). Spindle Checkpoint Silencing Requires Association of PP1 to Both Spc7 and Kinesin-8 Motors. *Developmental Cell*, 20(6):739–750.
25

# List of Figures

1        Overview. (A) Schematics of fission-yeast mitosis and kinesin-8 assembly states: Klp5/6 heterodimer, Klp5/5 homodimer, and Klp6 monomer. (B) Live-cell images of wild-type and kinesin8 deletion mutant cells with GFP-tagged spindle-pole bodies and kinetochore (green) as well as mCherry-tagged tubulin (red), illustrating the longer mitotic spindle length in cells with kinesin 8 deleted. (C) Fixed-cell images with GFP-tagged SPBs and KC (green), mCherry-tagged tubulin (red), and DAPI stained DNA (blue). Left, cell containing a lost KC, spindle, and polar MT bundle. Right, cell containing a KC near or attached to a polar MT bundle. (D-F) Image sequences; arrowhead indicates cen2 KC marker. (D) Wild-type cell illustrating lost KC reeling in, biorientation, segregation, and anaphase spindle elongation. See supplemental movie 1. (E) $5^+6\Delta$ cell illustrating lost KC reeling in, biorientation, spindle length fluctuations, segregation with lagging chromosome, and anaphase spindle elongation. See supplemental movie 2. (F) $5\Delta6^+$ cell illustrating KC reeling in, pushing, and biorientation. See supplemental movie 3. Scale bars are 1 $\mu$m.

2        Kinetochore movements after microtubule repolymerization. (A-I) Schematics and images. Green arrowheads indicate KC marker. (A-C) Schematic and image sequences illustrating reeling in of a lost KC to the SPB. In panel B, GFP tagged the SPBs and cen2 KC marker in a wild-type cell; successive images are 15 sec apart. In panel C, mCherry additionally tagged MTs in a $5^+6\Delta$ cell; successive images are 25 sec apart. See supplemental movie 4. (D-F) Schematic and image sequences illustrating KC pushing. In panel E, GFP tagged the SPBs and cen2 KC marker in a $5\Delta6\Delta$ cell; successive images are 2.5 min apart. See supplemental movie 5. In panel F, mCherry additionally tagged MTs in a $5\Delta6^+$ cell; successive images are 1.16 min apart. See supplemental movie 6. (G-H) Schematic and image sequences illustrating hovering of the KC near the SPB in $5\Delta6\Delta$ cell; successive images are 1.25 min apart. See supplemental movie 7. (I) Nuclear envelope deformation in $5^+6\Delta$ cell with long polar MTs. Left, mCherry channel showing polar MTs. Center, GFP channel showing SPBs and soluble nuclear GFP illustrating the deformation of the nuclear envelope by the MTs. Right, merge. Scale bars are 1 $\mu$m in all images. (J) Fraction of cells of different strains exhibiting pushing and hovering, which was never observed in wild-type cells. Error bars are standard deviation of a binomial distribution. (K) Fraction of cells of different strains for which a lost KC had reeled in to the SPB by different times after temperature shift. Error bars are standard deviation of a binomial distribution. (L) Length distribution of mitotic polar MT bundles in different strains with exponential fits. In lower two panels, the first two bins were excluded from the fit. Errors in characteristic length are uncertainties from fit. (M) Table quantifying reeling time, initial lost KC fraction, reeling speed, and length change. (N) Examples of 3D SPB-KC distance versus time illustrating reeling and pushing movements. Solid lines indicate reeling events with fits to determine speed. (O) Histograms of reeling speeds in different strains.

3        Kinetochore pushing movements and tripolar mitotic spindles. Schematics and images of cells containing SPBs tagged with sid4-mCherry SPB marker and microtubules tagged with mCherry-atb2 under a weak promoter (red, top), kinetochores tagged with mis6-GFP and mis12-GFP (green, middle), and merged images (bottom), all in the $5\Delta6^+$ background. (A) Chromosome pushing movements showing KC (arrowhead) near the end of a polar MT. See supplemental movies 9 and 10. (B) Tripolar mitotic spindle showing KC (arrowhead) colocalized with two bright and one dim SPB. See supplemental movie 11. (C) Chromosome pushing movements and tripolar spindle formation in the same cell. Initial images show spindle with polar MT extending up and right. At 4:30 the upper left SPB appears to split, forming a tripolar spindle that persists until the last frame. Also at 4:30 a KC (arrowhead) begins moving up and right along the polar MT, then reels back in to the SPB in the last two frames. See supplemental movie 12. Scale bars are 1 $\mu$m in all images. (D) Comparison of KC brightness per cell in cells with apparently normal KC dynamics (left, 268 images from 8 cells) and aberrant dynamics (right, 513 images from 13 cells).

4        Mitotic kinesin-8 localization. Images of cells containing fluorescently tagged Klp5/6 with and without the partner protein. (A) Fixed cell images with mCherry-tagged $\alpha$-tubulin (red), GFP-tagged Klp5/6 (green), DAPI-stained DNA (blue), and merged image (right). Images were taken with identical exposure conditions for each cell and brightness and contrast settings for the green channel. (B) Live cell images with mCherry-tagged $\alpha$-tubulin (red), GFP-tagged Klp5/6 (green), and merged image (right). Images were taken with identical exposure conditions for each cell and brightness and contrast settings for the green channel. (C, D) Live cell images with mCherry-tagged $\alpha$-tubulin (red) GFP-tagged Klp5/6 (green), and merged image (right), acquired with a newer EMCCD camera and displayed with brightness and contrast adjustment. (E, F) Live cell time-lapse images with mCherry-tagged $\alpha$-tubulin (red), GFP-tagged Klp5/6 (green), and merged image (right) acquired with a newer EMCCD camera and displayed with brightness and contrast adjustment. Sequential images are separated by 5 minutes. Scale bars are 1 $\mu$m in all images. (G) Summary statistics of mitotic Klp5/6 localization in the absence of the partner motor. Data indicate fraction of structures with GFP labeling. Klp5-GFP spindle and near-SPB localization data are from 50 cells, and polar MT localization data are from 17 cells with observable polar MTs. Klp6-GFP localization data are from 50 cells, and polar MT localization data are from 15 cells with observable polar MTs. Error bars are standard deviation of a binomial distribution. Differences between $5^{GFP}6\Delta$ and $5\Delta6^{GFP}$ cells were statistically significant for all three types of localization studied (see text).

5      Biorientation. (A-B) Schematic and image sequences of biorientation in wild-type and kinesin-8 deletion cells; successive images are 1 min apart. See supplemental movies 13-14. Scale bars are 1 $\mu$m and arrowheads indicate KC marker. (C) Fraction of cells of different strains for which a KC hasbioriented by different times after temperature shift. Error bars are standard deviations of a binomial distribution. (D) Quantification of time to biorientation. (E) Examples of 3D SPB-KC distance and sister KC separation versus time showing biorientation, KC movement along the spindle, and sister KC breathing. (F-G) Quantification of KC position along the spindle. Note different $y$-axis scales for different strains. (F) Histograms of absolute distance of bioriented KCs from reeling SPB. (G) Same data as in (F) plotted as fractional position between the two SPBs. (H) Histograms of sister KC separation. Note different $y$-axis scales for different strains and that height of the large peak at zero separation is not shown.

6      Spindle length dynamics. (A) Schematic of spindle length fluctuations with green dots representing SPBs, red bars representing the mitotic spindle, and black lines representing the nuclear envelope. (B) Examples of mitotic spindle length (represented by 3D SPB-SPB distance) illustrating short, stable spindle lengths in wild-type cells and longer, fluctuating spindle lengths in kinesin-8 deletion cells. (C) Examples of spindle length versus time with chromosome segregation events indicated by large circles. (D) Temporal autocorrelation function of fluctuations in spindle length.

7      Chromosome segregation. (A-B) Schematic and image sequence of a lagging chromosome; successive images are 15 sec apart. See supplemental movie 15. Scale bar is 1 $\mu$m and arrowheads indicate KC marker. (C) Fraction of cells of different strains exhibiting lagging chromosomes. Error bars are standard deviations of a binomial distribution. (D) Fraction of cells of different strains for which chromosome segregation has occurred by different times after temperature shift. Error bars are standard deviations of a binomial distribution. (E) Examples of quantification of 3D SPB-KC distance of each sister KC from the SPB to which it is segregated. Solid lines indicate segregation events with fits to determine speed. (F) Histogram of segregation speeds in different strains. (G) Quantification of segregation events.

8      Anaphase spindle elongation. (A-D) Schematics and images of anaphase spindle elongation and nuclear-envelope deformation in $5^+6\Delta$ cells. Left, spindle (red); center, SPBs and soluble nuclear GFP (green); right, merged image. (A,C) nuclear envelope deformation into a peanut-like shape. (B,D) Nuclear envelope deformation by extension of a narrow tether. Scale bars are 1 $\mu$m. (E) Example of spindle length (represented by 3D SPB-SPB distance) versus time during anaphase spindle elongation. Solid line indicates elongation event with fit to determine speed. (F) Histogram of elongation speeds in different strains. (G) Quantification of elongation events.

9      Microtubule bundle model. (A-B) Schematic and key model features. (C) Wild-type model results, including experimental and model simulated traces (left), SPB-KC length distribution (center), and temporal autocorrelation of fluctuations in SPB-KC length (right). (D) $5^+6\Delta$ model results. (E)   $5\Delta$ model results.

10     Spindle force balance model. (A-B) Schematic and key model features. (C, E, G) Experimental and simulated spindle length versus time. (D, F, H) Distribution of KC positions along the spindle.(C-D) Wild-type model results. (E-F) $5^+6\Delta$ model results. (E) $5\Delta$ model results.

11    Conceptual model of Klp5/6 mitotic effects.

## List of Tables

1     Effects of cold treatment on cell viability. Growth rates of strains of each genotype were measured before and after induction of lost KCs by cold treatment. Growth rates increased similarly for all strains.

2     Results of automated analysis of spindle length and length change events.

3      Model parameters that were different in parameter sets corresponding to different strains. Parameters were determined through comparison to experimental data except as noted. *Functional form of Tischer et al. (2009): the rate of increase of catastrophe frequency with length is 37.5% of the bare catastrophe frequency. †Akiyoshi et al. (2010).

| | |
|---|---|
| 4 | Strains used in this study. |
| 5 | Pushing bundle model fixed parameters. |
| 6 | Spindle force balance model fixed parameters. |

| Strain | Doubling time (hours, fit± err) | Doubling time after cold treatment (hours, fit± err) | Increase in doubling time (value±err) |
|---|---|---|---|
| WT | 2.49±0.10 | 3.58±0.25 | 1.44±0.12 |
| 5⁺6Δ | 2.65±0.05 | 3.01±0.05 | 1.14±0.03 |
| 5Δ6⁺ | 2.51±0.11 | 3.12±0.15 | 1.24±0.08 |
| 5Δ6Δ | 2.56±0.05 | 3.38±0.18 | 1.32±0.07 |

Table 1: Effects of cold treatment on cell viability. Growth rates of strains of each genotype were measured before and after induction of lost KCs by cold treatment. Growth rates increased similarly for all strains.

| Strain | Spindle length (μm, mean ± SEM) | Length change shortening (μm, mean ± SEM) | Length change lengthening (μm, mean ± SEM) | Duration shortening (min, mean ± SEM) | Duration lengthening (min, mean ± SEM) | Speed shortening (μm/min, mean ± SEM) | Speed lengthening (μm/min, mean ± SEM) |
|---|---|---|---|---|---|---|---|
| WT | 1.79±.12 (N= 16) | 0.40±0.06 (N=20) | 0.60±0.06 (N=54) | 2.01±0.08 | 2.63±0.13 | -0.22±0.04 | 0.23±0.02 |
| 5⁺6Δ | 4.45±0.22 (N= 33) | 1.02±0.20 (N=37) | 1.56±0.17 (N=98) | 2.78±0.20 | 5.05±0.37 | -0.35±0.06 | 0.29±0.01 |
| 5Δ6⁺ | 3.17±0.28 (N=17) | 1.30±0.31 (N=28) | 1.28±0.18 (N=49) | 3.18±0.40 | 4.21±0.47 | -0.40±0.06 | 0.30±0.02 |
| 5Δ6Δ | 3.46±0.18 (N=29) | 0.74±0.12 (N=38) | 1.80±0.25 (N=56) | 2.36±0.14 | 5.07±0.57 | -0.33±0.05 | 0.37±0.02 |

Table 2: Results of automated analysis of spindle length and length change events.

|  | WT | 5⁺6Δ | 5Δ |
|---|---|---|---|
| **Bundle and spindle model** | | | |
| Growth speed ($\mu$m/min) | 0.5 | 0.6 | 0.7 |
| Shrinking speed ($\mu$m/min) | 1.5 | 1.5 | 0.5 |
| Catastrophe frequency (/min) | $0.5 + 0.1875\, L_{MT}/\mu$m* | 0.15 (attached) 0.015 (unattached) | 0.015 |
| Rescue frequency (/min) | 1.0 | 0.1 (attached) 0.04 (unattached) | 0.04 |
| Unbinding rate of growing MT (/min) | 10⁻⁴ | $6 \times 10^{-3}$† | $6 \times 10^{-3}$† |
| Unbinding rate of shrinking MT (/min) | 10⁻³ | $4.8 \times 10^{-2}$† | $4.8 \times 10^{-2}$† |
| **Bundle model only** | | | |
| Attachment frequency (/min) | 0.5 | 2.0 | 1.0 |
| Nucleation frequency (/min) | 1.0 | 0.1 | 1.0 |
| Initial MT number | 3 | 6 | 10 |
| **Spindle model only** | | | |
| Sliding force (pN) | 13 | 14 | 4 |

Table 3: Model parameters that were different in parameter sets corresponding to different strains. Parameters were determined through comparison to experimental data except as noted. *Functional form of Tischer et al. (2009): the rate of increase of catastrophe frequency with length is 37.5% of the bare catastrophe frequency. †Akiyoshi et al. (2010).

| | Value | Reference |
|---|---|---|
| MT growth force constant (pN/$\mu$m) | 8.4 | Akiyoshi et al. (2010) |
| MT shrinking force constant (pN/$\mu$m) | 3.0 | Akiyoshi et al. (2010) |
| MT catastrophe force constant (pN/$\mu$m) | 2.3 | Akiyoshi et al. (2010) |
| MT rescue force constant (pN/$\mu$m) | 6.4 | Akiyoshi et al. (2010) |
| MT unbinding force constant (pN/$\mu$m) | 4.0 | Akiyoshi et al. (2010) |
| MT polymerization stall force (pN) | 12 | Chosen > 5 based on van Doorn et al. (2000) |
| Nuclear envelope deformation force (pN) | 17 (bundle) 14.5 (spindle) | For a tether of radius 50 nm (bundle) or 100 nm (spindle), the tether extension force = $2\pi R[(\sigma + \kappa/(2R^2)]$, with $\kappa = 2 \times 10^{-19}$ J and $\sigma = 1.3 \times 10^{-5}$ N/m, based on Lim et al. (2007) |
| Unbound KC diffusion coefficient ($\mu$m$^2$/min) | 0.084 | Kalinina et al. (2012) |
| MT-KC interaction distance (nm) | 100 (bundle) 10 (spindle) | Chosen |
| MT-KC linkage spring constant (pN/$\mu$m) | 40 | Gay et al. (2012) |

Table 5: Pushing bundle model fixed parameters.

| | Value | Reference |
|---|---|---|
| Number of KCs per half spindle | 3 | Ding et al. (1993) |
| Number of MTs per KC | 3 | Ding et al. (1993) |
| Nuclear radius ($\mu$m) | 1 onset, 1.2 max | Our observations |
| Maximum MT shrinking speed ($\mu$m/min) | 30 | Sagolla et al. (2003) |
| Inter-KC linkage spring constant (pN/$\mu$m) | 5 | Reduced slightly from Gay et al. (2012) |
| Sister KC overlap spring constant (pN/$\mu$m) | 20 | Chosen larger than linkage spring constant |
| Sister KC offset (nm) | 50 | Chosen |
| MT-SPB overlap spring constant (pN/$\mu$m) | 40 | Chosen |
| SPB friction coefficient (pN min/$\mu$m) | 25 | Gay et al. (2012) |

Table 6: Spindle force balance model fixed parameters.

| Strain | Genotype | Source |
|---|---|---|
| McI730 | nda3-KM311, cen2::kan$^r$-ura4$^+$-lacOp his7$^+$::lacI-GFP, nmt1-GFP-pcp1$^+$::kan$^r$, mcherry-atb2:natMX6, leu1-32, ura-D18, h$^-$ | This study |
| McI718 | nda3-KM311, cen2::kan$^r$-ura4$^+$-lacOp his7$^+$::lacI-GFP, nmt1-GFP-pcp1$^+$::kan$^r$, Klp5D::ura4$^+$, klp6D::his3$^+$, ade6-M216, ura4-D18, leu1-32, h$^+$ | This study |
| McI731 | nda3-KM311, cen2::kan$^r$-ura4$^+$-lacOp his7$^+$::lacI-GFP, nmt1-GFP-pcp1$^+$::kan$^r$, Klp5D::ura4$^+$, klp6D::his3$^+$, ade6-M216, ura4-D18, leu1-32, h$^+$ | This study |
| McI765 | nda3-KM311, cen2::kan$^r$-ura4$^+$-lacOp his7$^+$::lacI-GFP, nmt1-GFP-pcp1$^+$::kan$^r$, Klp5D::ura4$^+$, his3-D1, ura4-D18, leu1-32, h$^+$ | This study |
| McI770 | nda3-KM311, cen2::kan$^r$-ura4$^+$-lacOp his7$^+$::lacI-GFP, nmt1-GFP-pcp1$^+$::kan$^r$, Klp5D::ura4$^+$, ade6-M216, his3-D1, ura4-D18, leu1-32, h$^+$ | This study |
| McI773 | nda3-KM311, cen2::kan$^r$-ura4$^+$-lacOp his7$^+$::lacI-GFP, nmt1-GFP-pcp1$^+$::kan$^r$, Klp5D::ura4$^+$, mcherry-atb2:natMX6, ura4-D18, leu1-32, h90 | This study |
| McI721 | nda3-KM311, cen2::kan$^r$-ura4$^+$-lacOp his7$^+$::lacI-GFP, nmt1-GFP-pcp1$^+$::kan$^r$, klp6D::his3, ade6-M216, ura4-D18, leu1-32, h$^-$ | This study |
| McI746 | nda3-KM311, cen2::kan$^r$-ura4$^+$-lacOp his7$^+$::lacI-GFP, nmt1-GFP-pcp1$^+$::kan$^r$, klp6D::his3$^+$, mcherry-atb2:natMX6, ade6-M216, leu1-32, h$^-$ | This study |
| McI748 | nda3-KM311, cen2::kan$^r$-ura4$^+$-lacOp his7$^+$::lacI-GFP, nmt1-GFP-pcp1$^+$::kan$^r$, klp6D::his3$^+$, mcherry-atb2:natMX6, leu1-32, h$^-$ | This study |
| McI726 | nda3-KM311, cen2::kan$^r$-ura4$^+$-lacOp his7$^+$::lacI-GFP, nmt1-GFP-pcp1$^+$::kan$^r$, klp6D::his3$^+$, klp2D::ura4$^+$, ade6-M216, ura4-D18, leu1-32, h$^-$ | This study |
| McI833 | nda3-KM311, mis6-GFP:kanMX6, mis12-GFP::leu1$^+$, sid4-mcherry:natMX6, klp5D::ura4$^+$, pnda3-mcherry-atb2:aur1$^r$, h$^-$ | This study |
| McI837 | nda3-KM311, mis6-GFP:kanMX6, mis12-GFP::leu1$^+$, sid4-mcherry:natMX6, pnda3-mcherry-atb2:aur1$^r$, ura4-D18, ade6-M216, h$^+$ | This study |
| McI847 | nda3-KM311, mis6-GFP:kanMX6, mis12-GFP::leu1$^+$, sid4-mcherry:natMX6, klp6D::ura4$^+$, pnda3-mcherry-atb2:aur1$^r$, ade6-M216, h$^+$ | This study |
| Original strains | | |
| K39 | nda3-KM311, mis12-GFP::leu1$^+$, klp5D:: ura4$^+$, leu1-32, ura4-D18, h$^-$ | R. McIntosh and E. Grishchuk |
| K41 | nda3-KM311, mis12-GFP::leu1$^+$, klp6D:: ura4$^+$, leu1-32, ura4-D18, h$^-$ | R. McIntosh and E. Grishchuk |
| K64 | nda3-KM311, cen2::kan$^r$-ura4$^+$-lacOp his7$^+$::lacI-GFP, nmt1-GFP-pcp1$^+$::kan$^r$, leu1-32 ura4-D18, h$^-$ | R. McIntosh and E. Grishchuk |
| K70 | nda3-KM311, klp2D::ura4$^+$, cen2-GFP, pcp1-GFP, leu1-32, ura4-D18, h$^-$ | R. McIntosh and E. Grishchuk |
| McI397 | klp5D::ura4$^+$, klp6D::his3$^+$, ade6-M216, his3-D1, leu1-32, ura4-D18, h$^+$ | R. McIntosh |
| McI381 | klp5D::ura4$^+$, ade6-M216, his3-D1, leu1-32, ura4-D18, h$^+$ | R. McIntosh |
| McI485 | klp5:GFP:ura4$^+$, ade6-m210, his3-D1, leu1-32, ura4-D18, h$^+$ | R. McIntosh |
| McI486 | klp6:GFP:ura4$^+$, ade6-m210, his3-D1, leu1-32, ura4-D18, h$^-$ | R. McIntosh |
| McI487 | klp5D::ura4$^+$, klp6:GFP:ura4$^+$, ade6-M216, his3-D1, leu1-32, ura4-D18, h$^+$ | R. McIntosh |
| McI488 | klp6D::his3$^+$, klp5:GFP:ura4$^+$, ade6-M216, his3-D1, leu1-32, ura4-D18, h$^+$ | R. McIntosh |
| McI728 | z:adh15:mcherry-atb2:natMX6, leu1-32, ura4-D18, h$^+$ | Y. Watanabe |
| AR614 | klp5-GFP-kan, aur1-mcherry-atb2, leu1-32, ura4-D18 h$^-$ | T. Toda |
| AR615 | klp5-GFP-kan, klp6D::ura4$^+$, aur1-mcherry-atb2, leu1-32, ura4-D18 h$^-$ | T. Toda |
| AR616 | klp6-GFP-kan, aur1-mcherry-atb2, leu1-32, ura4-D18 h$^-$ | T. Toda |
| AR617 | klp6-GFP-kan, klp5D::ura4$^+$, aur1-mcherry-atb2, leu1-32, ura4-D18, his7 h$^-$ | T. Toda |
| JCF9907 | mis6-GFP:kanMX6, sid4-mcherry:natMX6, pnda3-mcherry-atb2:aur1$^r$, ade6-M216, h$^+$ | J. Cooper |

Table 4: Strains used in this study.

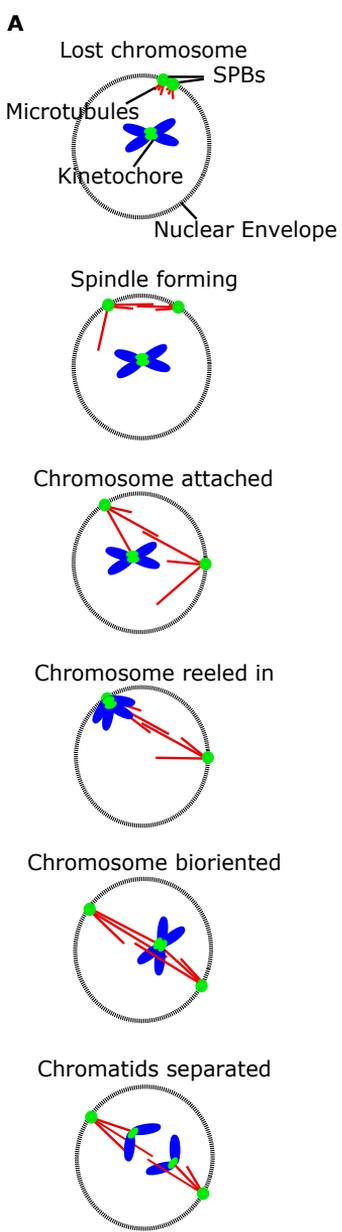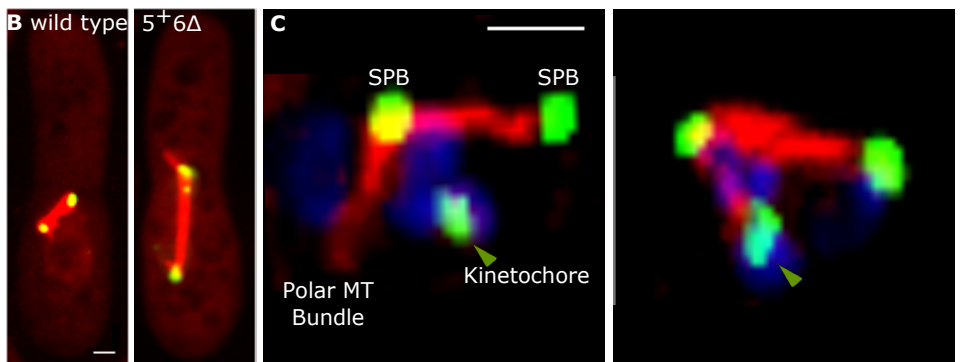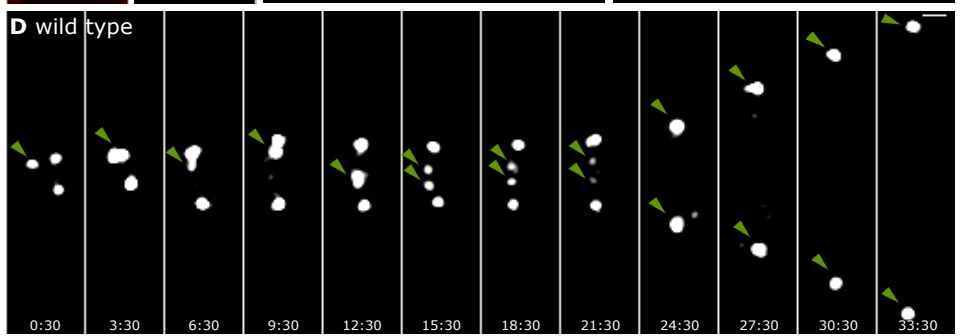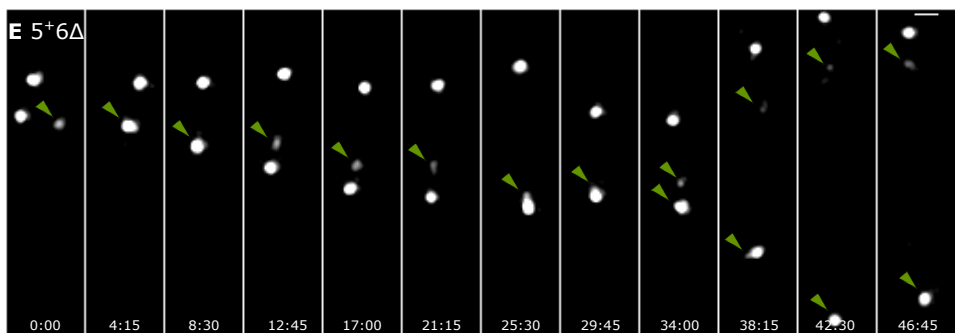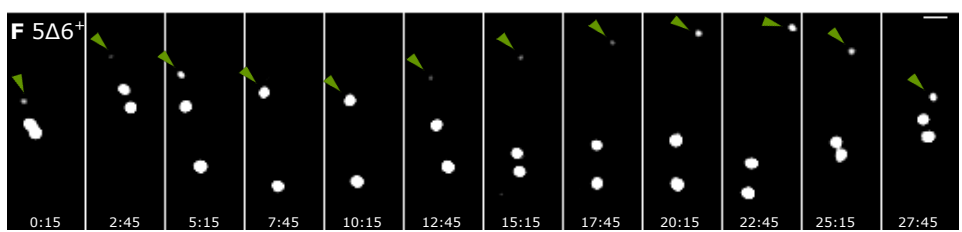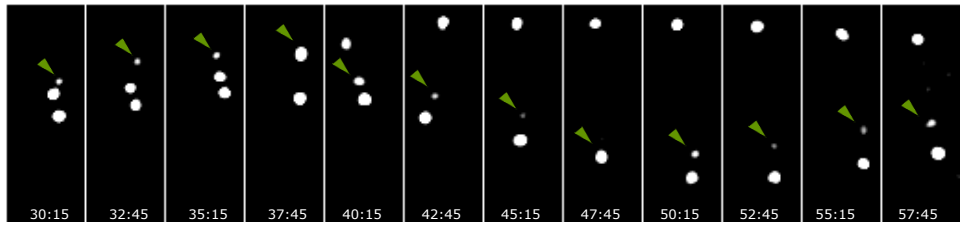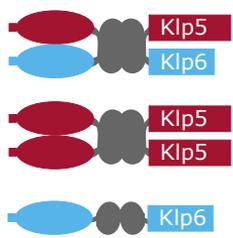

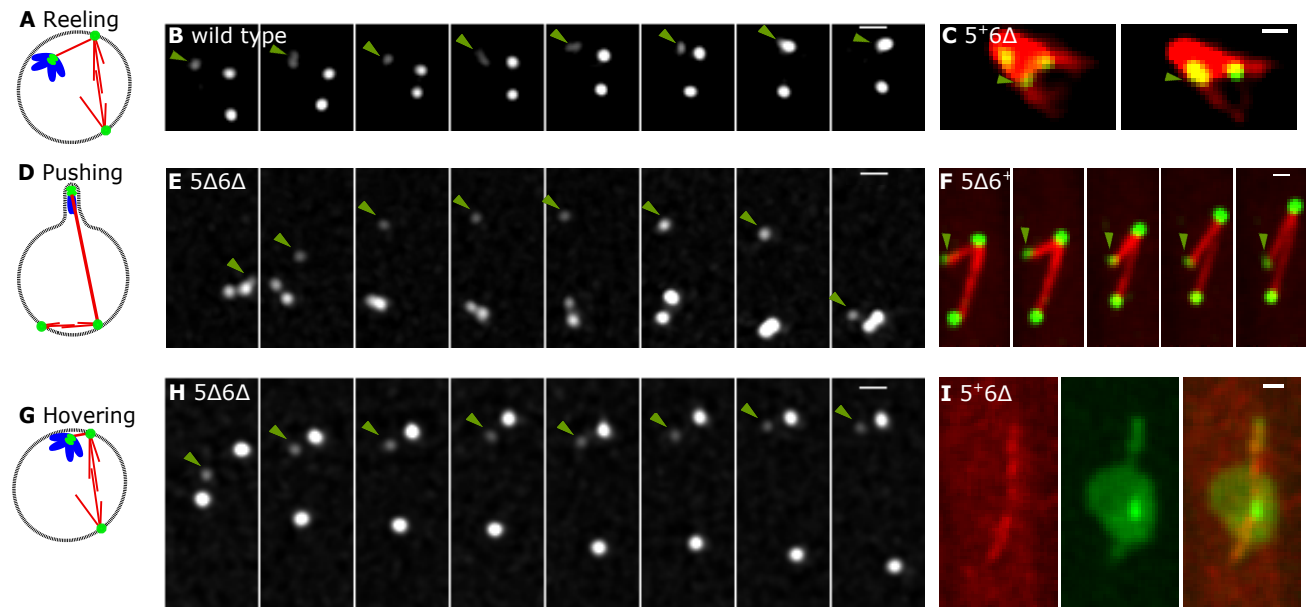
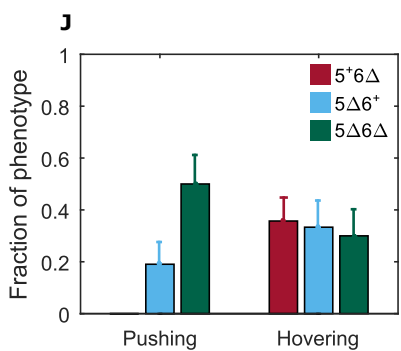
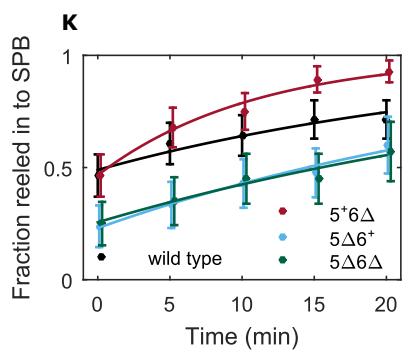
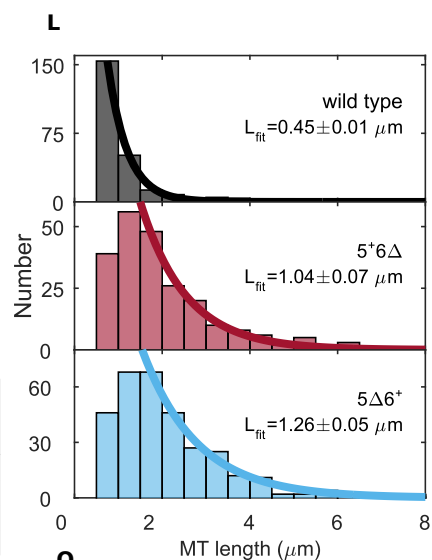
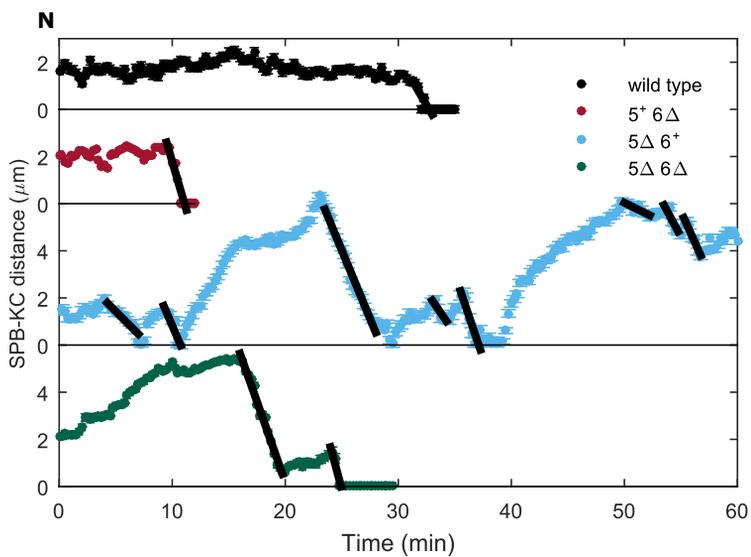
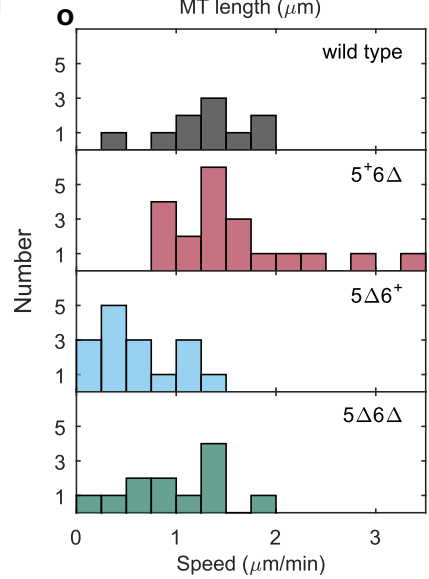

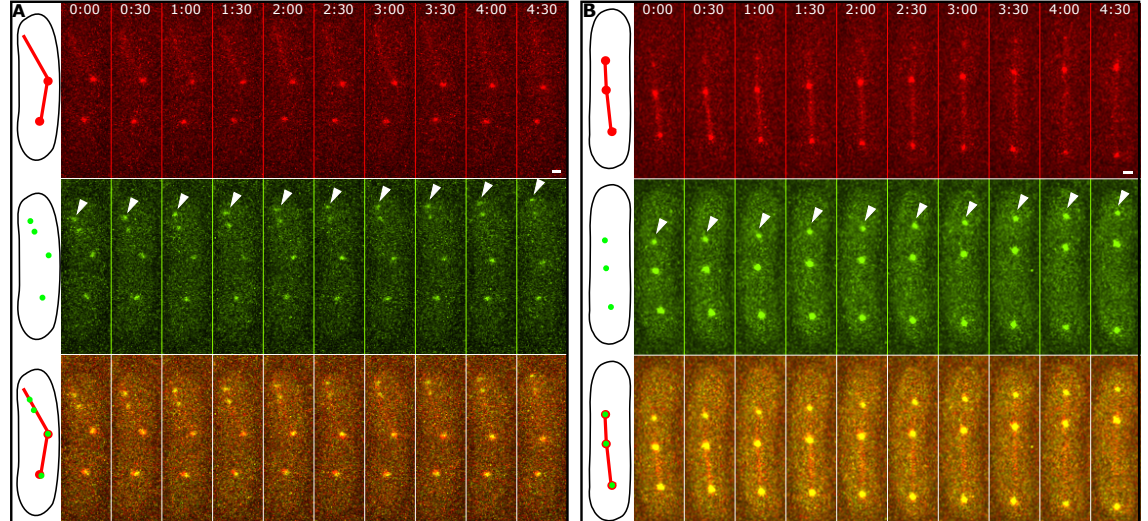
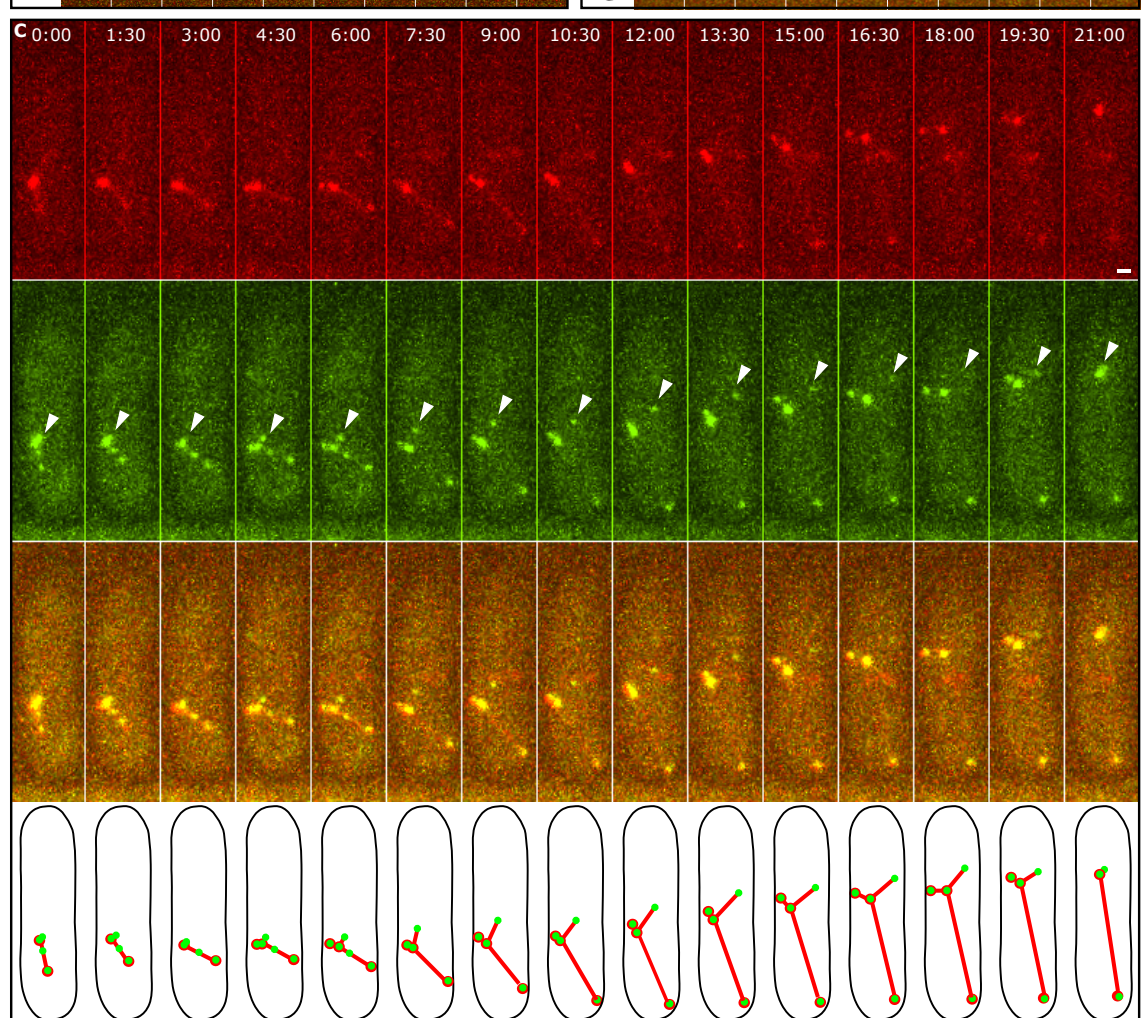
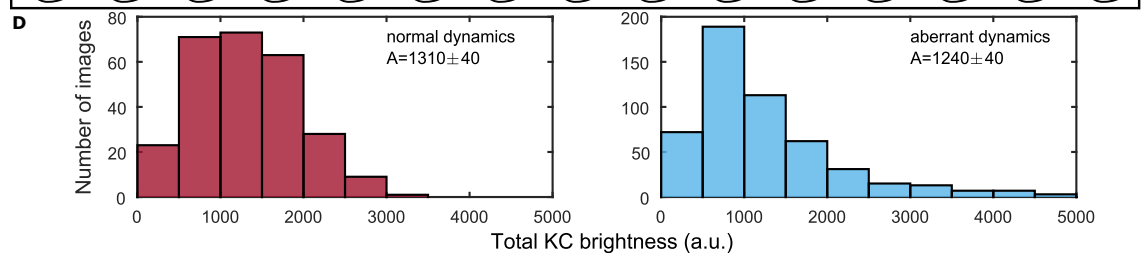

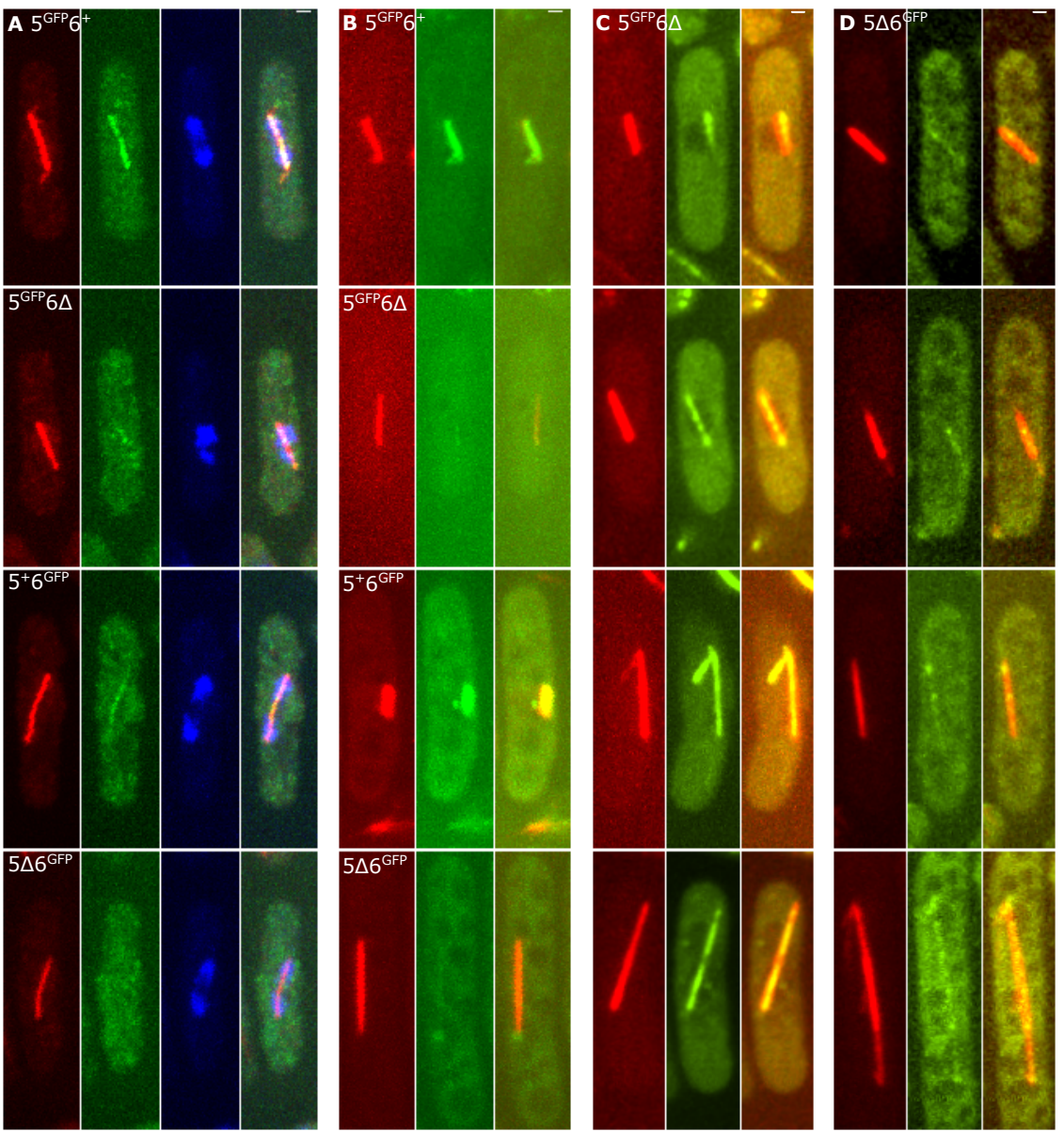
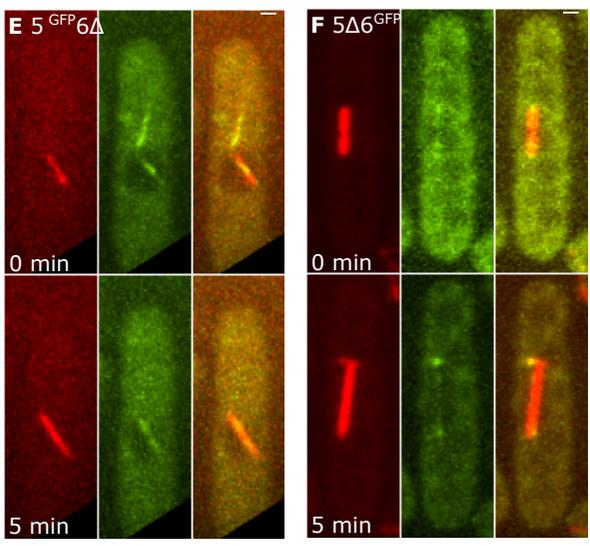
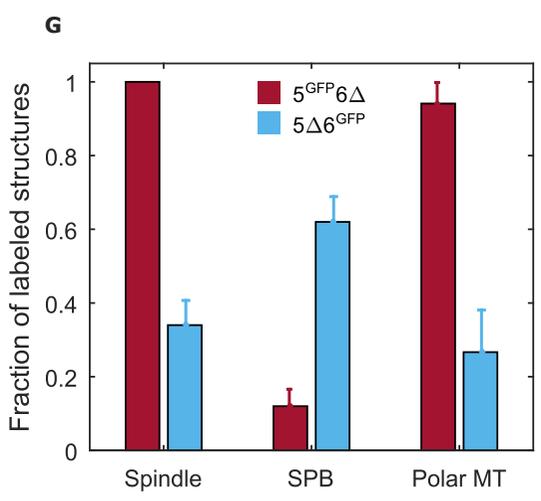

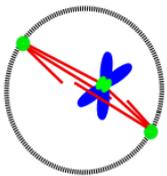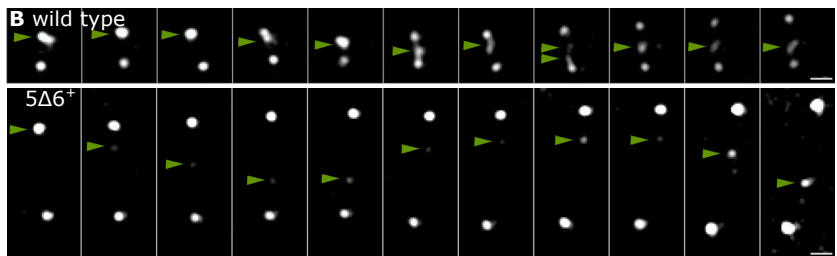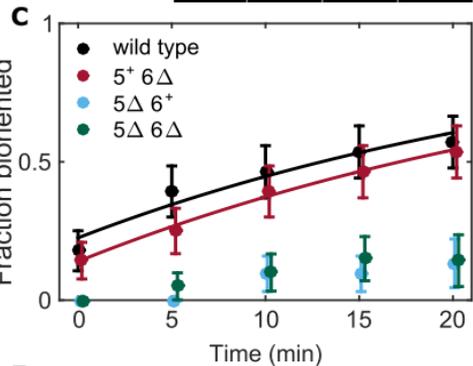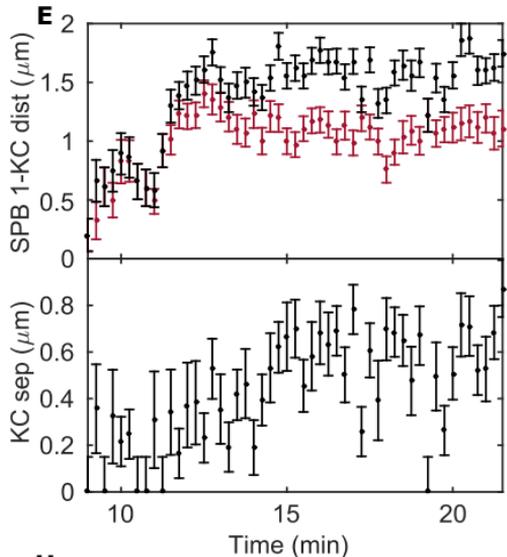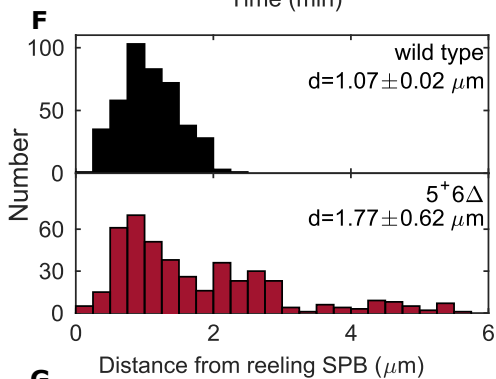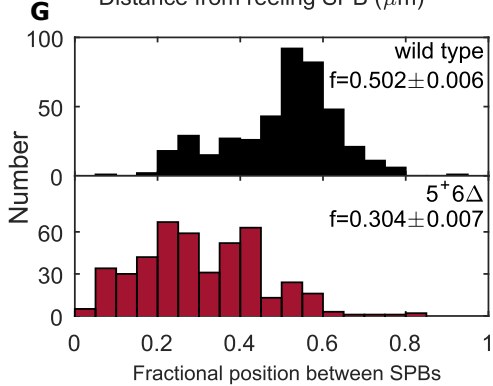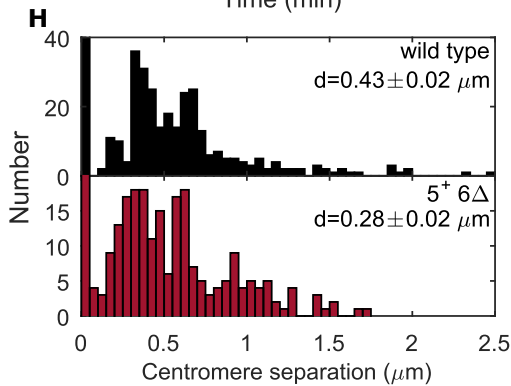

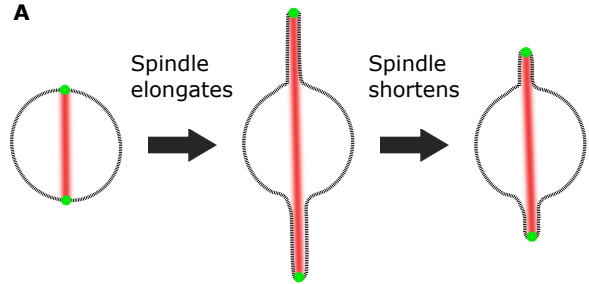
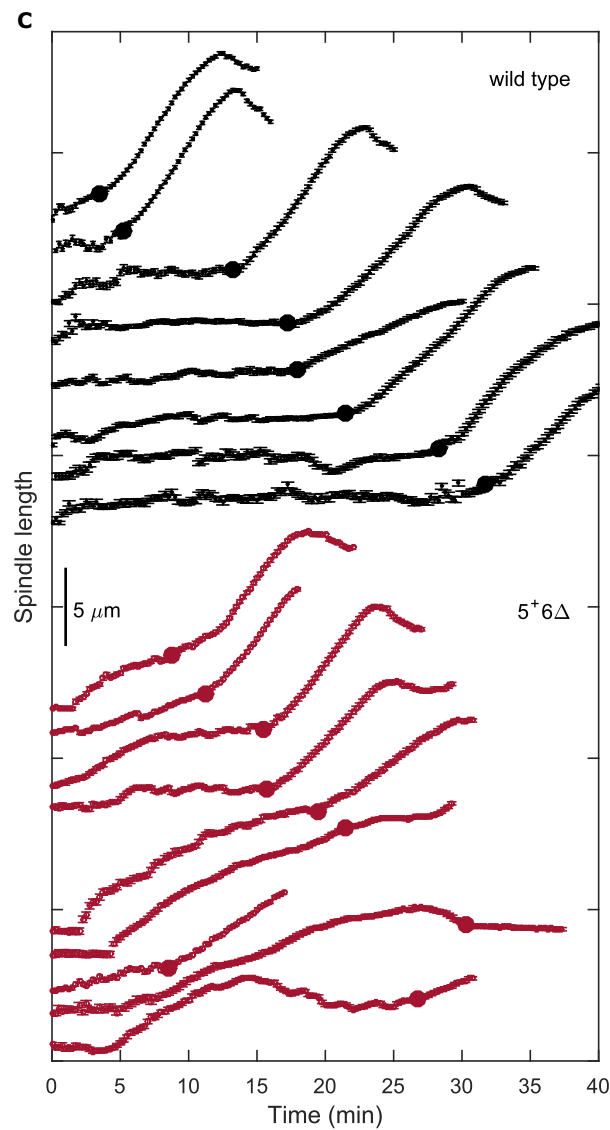
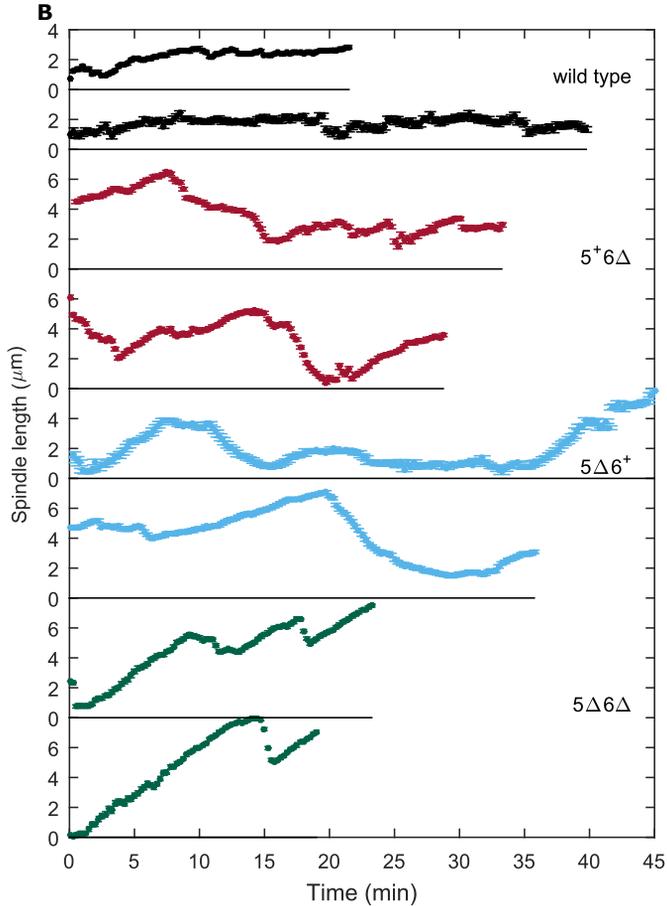
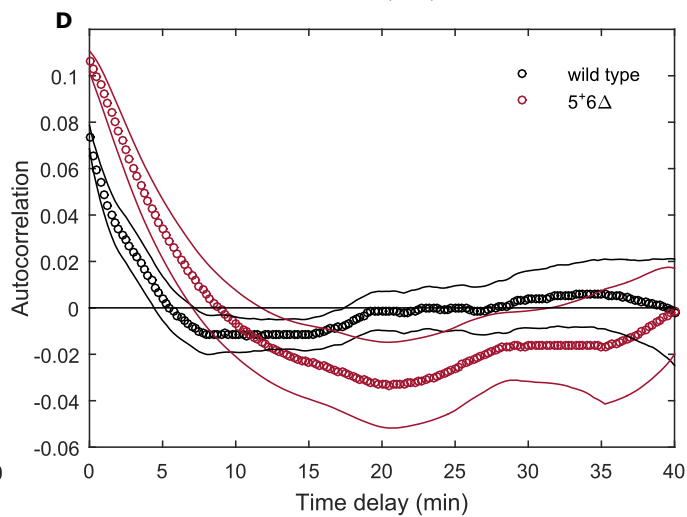

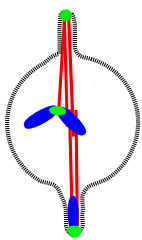
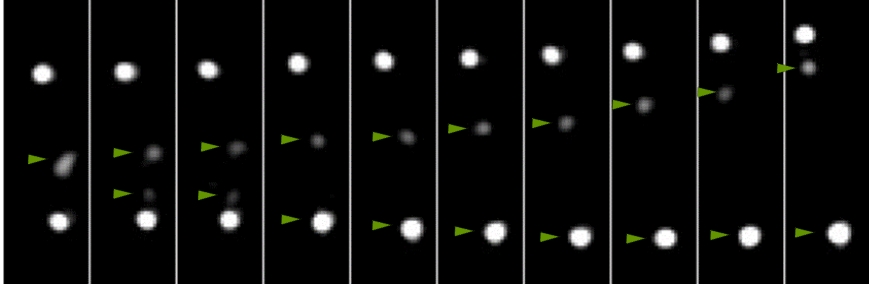
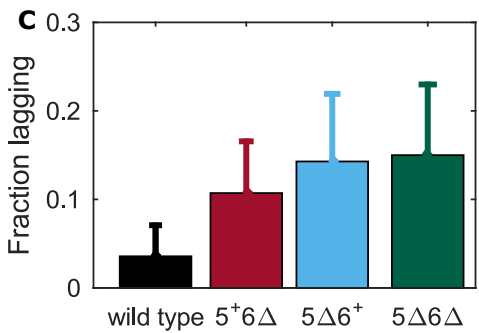
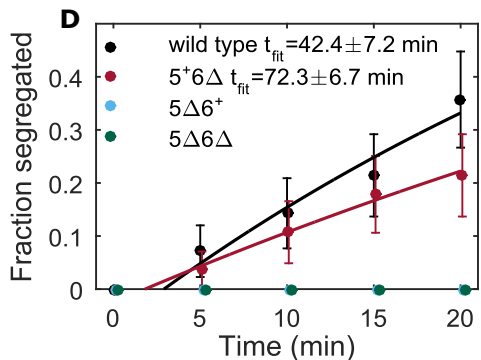
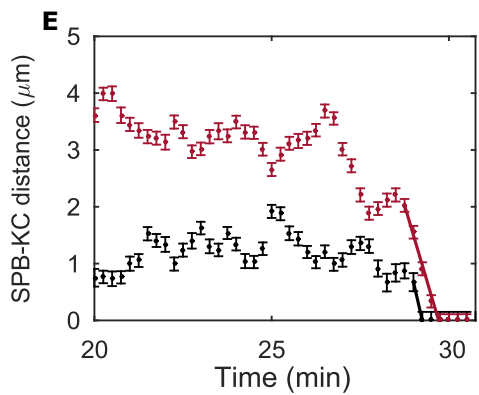
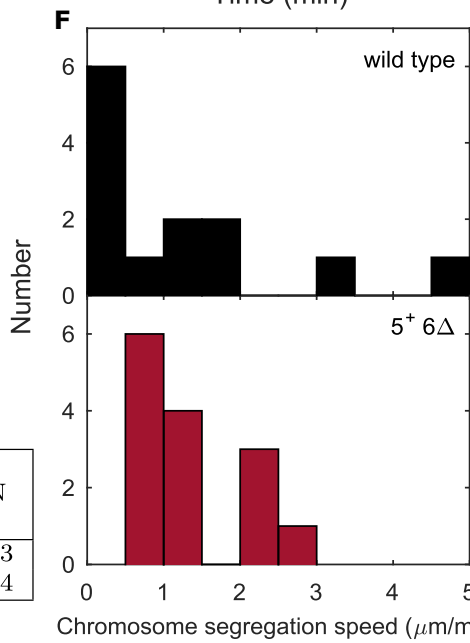

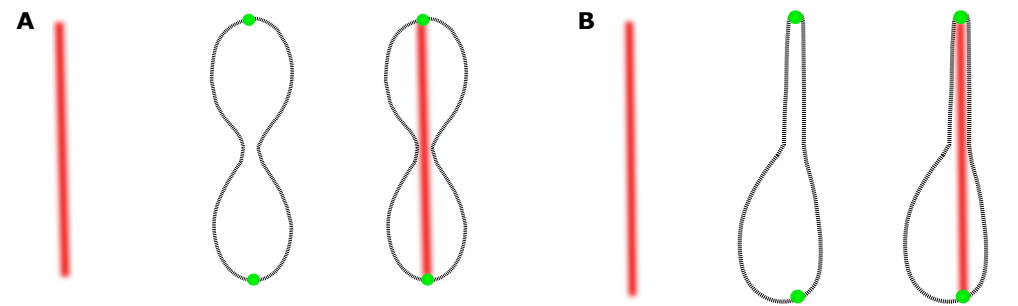

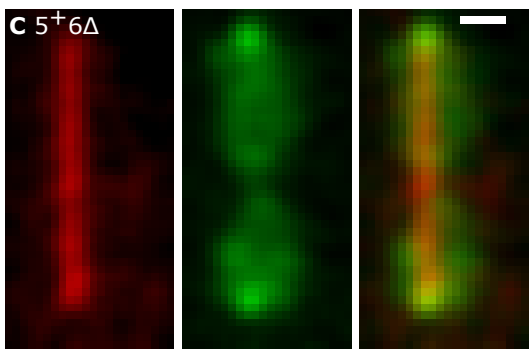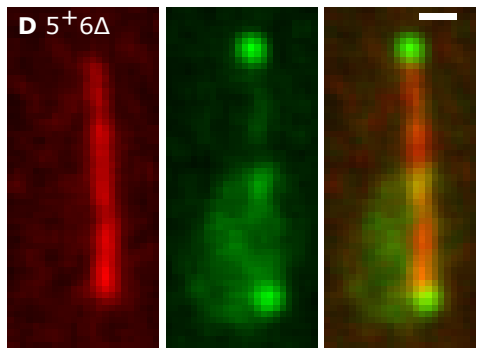

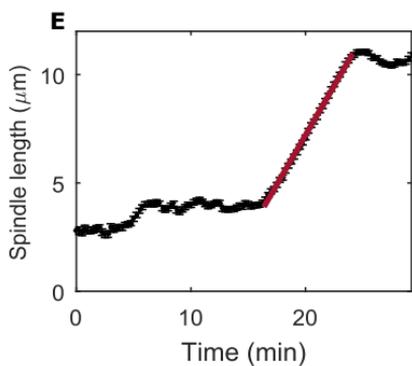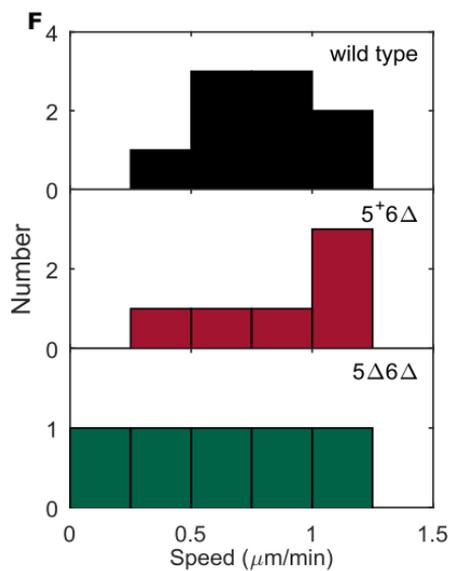

# Microtubule bundle model

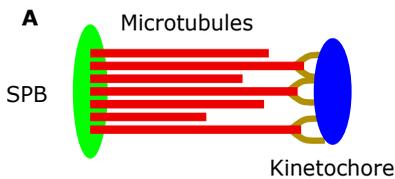

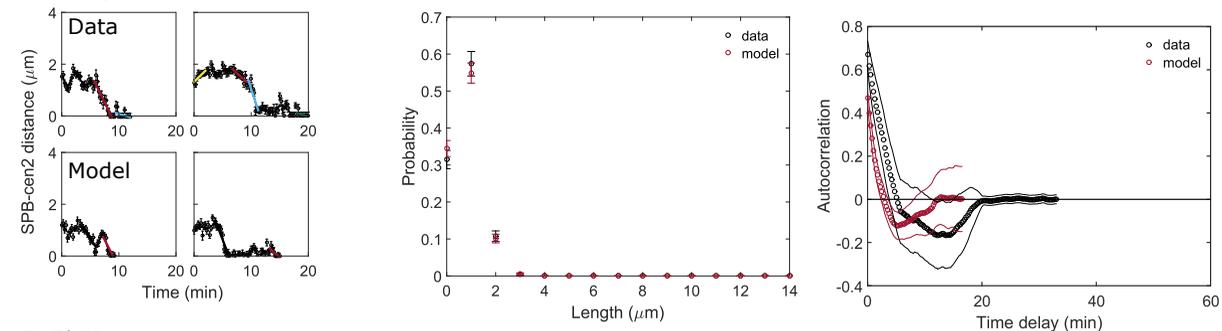

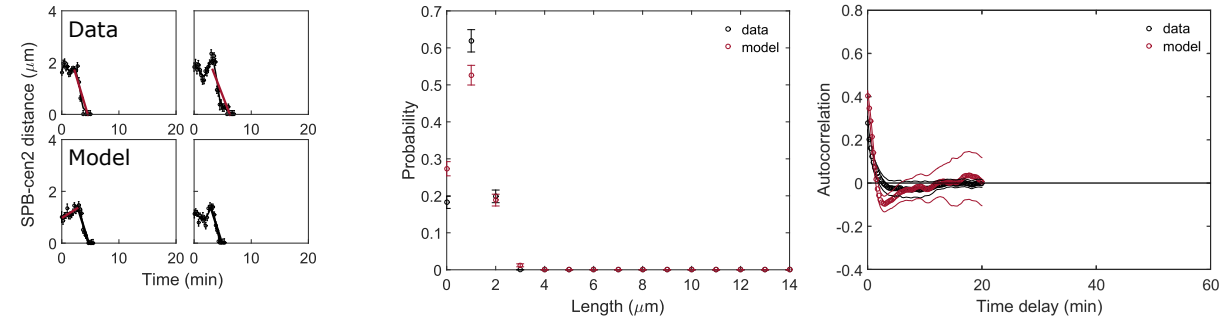

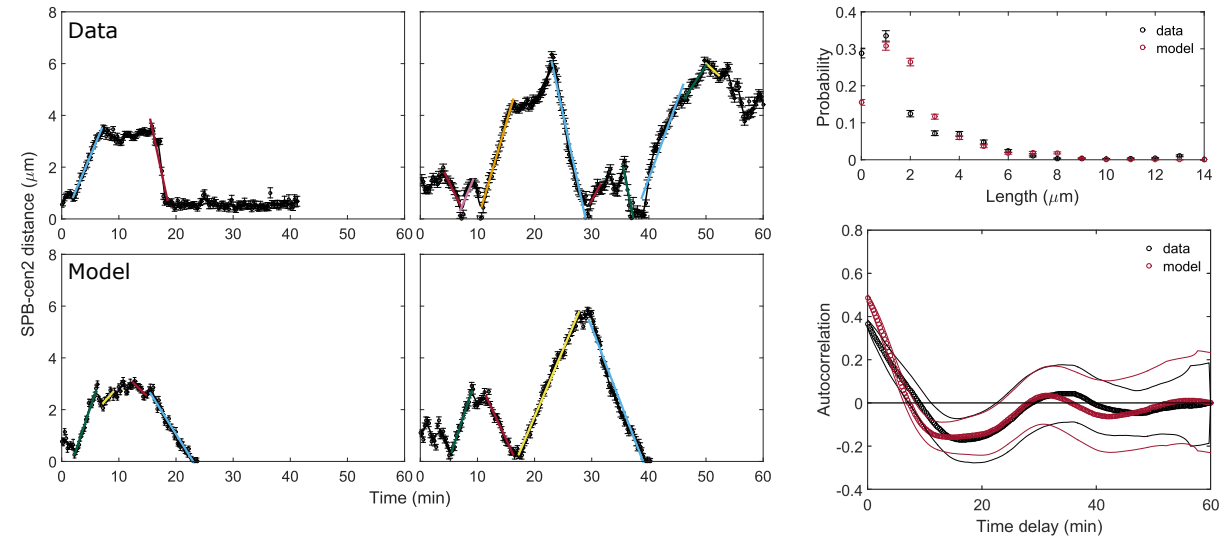

# Spindle force balance model

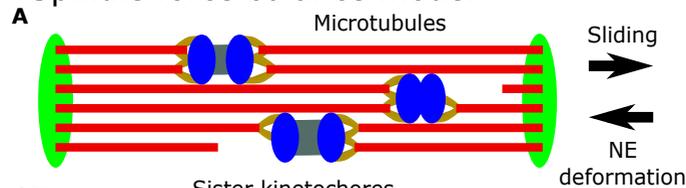
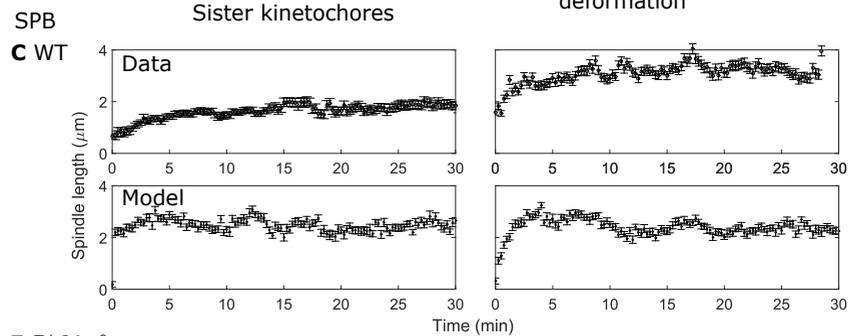
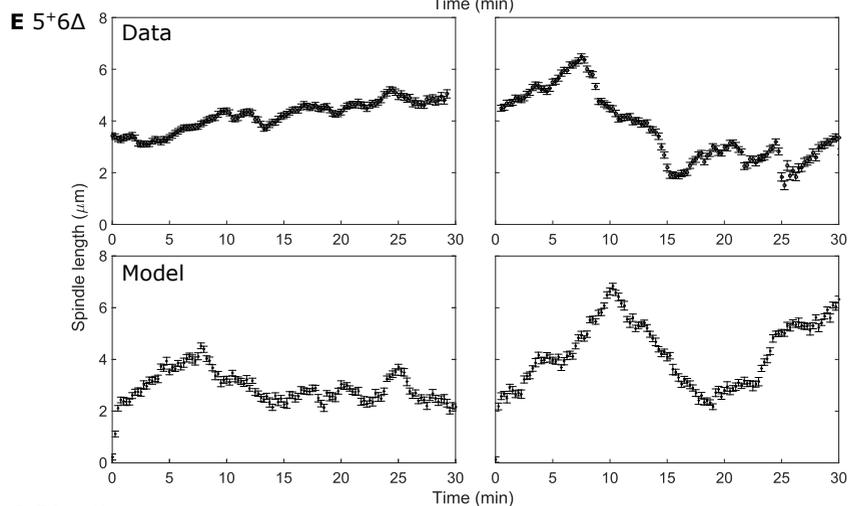
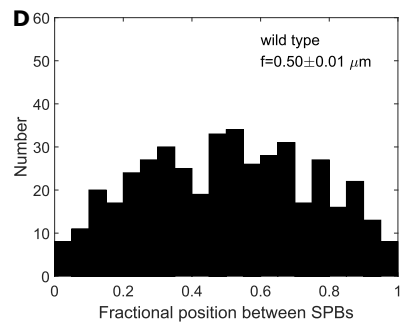
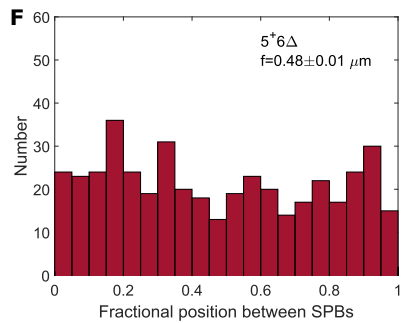
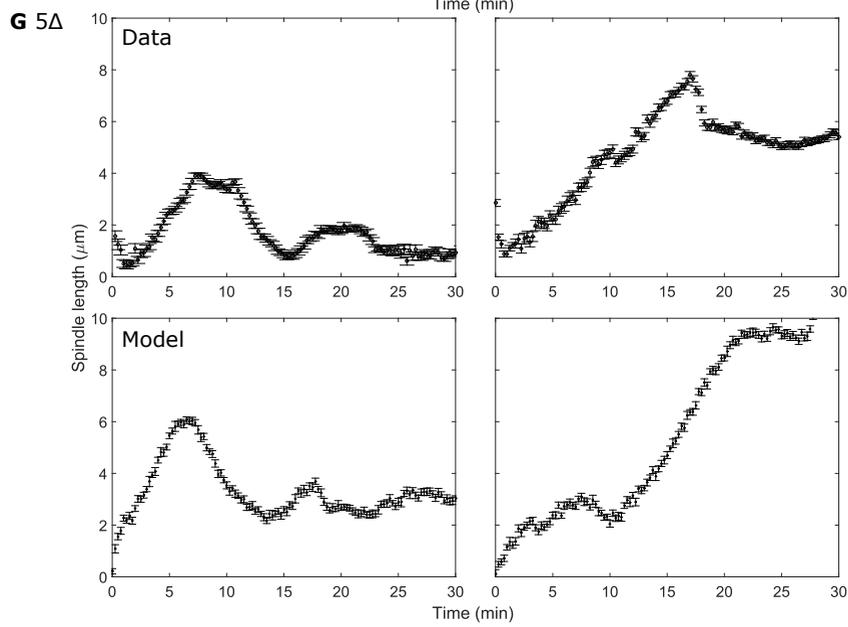
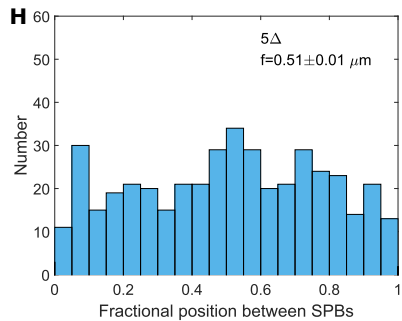

## A Wild type

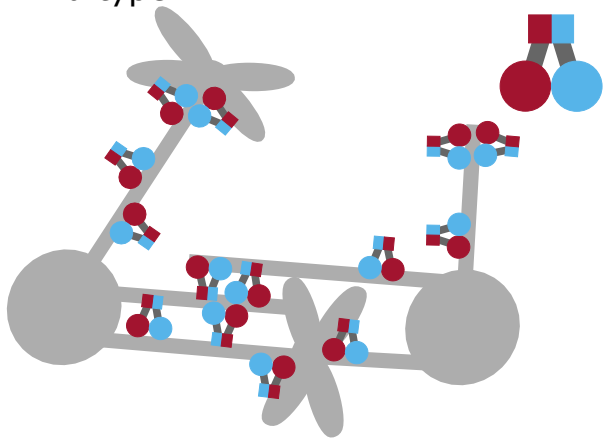

**Klp5/6 heterodimers**
**Mathematical model parameters**
Length-dependent catastrophe
High rescue frequency
Stable MT-KC attachments
High sliding force

**Mathematical model results**
Rapid reeling in of lost KCs
Stable spindle length
KC centering on spindle

## B $5^+6\Delta$

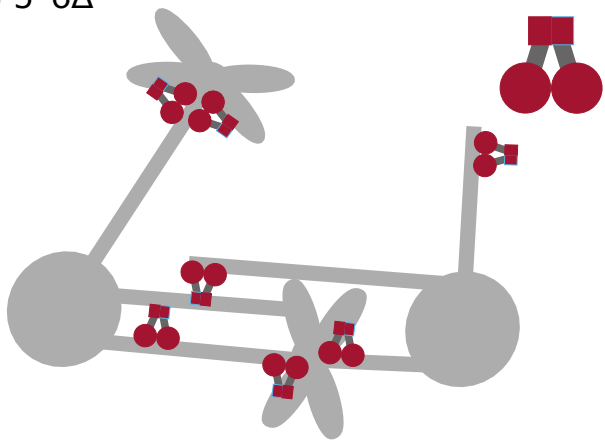

**Klp5 homodimers**
**Mathematical model parameters**
Length-independent catastrophe
Catastrophe and rescue frequency
 depend on attachment
Unstable MT-KC attachments
High sliding force

**Mathematical model results**
Rapid reeling in of lost KCs
Long spindle
Variable spindle stability
Loss of KC centering

## C $5\Delta6^+$

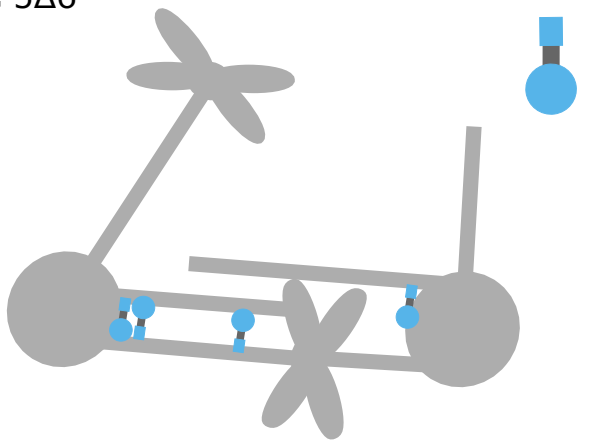

**Klp6 monomers**
**Mathematical model parameters**
Length-independent catastrophe
Low catastrophe and rescue frequencies
Unstable MT-KC attachments
Low sliding force

**Mathematical model results**
Pushing of KCs
Long, unstable spindle
Loss of KC centering